\documentclass[prd,tightenlines,nofootinbib,superscriptaddress]{revtex4}
\usepackage{amsfonts,amsmath,amssymb,amsthm,bbm,hyperref}

\usepackage{color,psfrag}
\usepackage[dvips]{graphicx}

\newtheorem{res}{Result}
\newtheorem{conj}{Conjecture}

\newcommand{\C}{{\mathbb C}}
\newcommand{\N}{{\mathbb N}}
\newcommand{\R}{{\mathbb R}}

\newcommand{\pP}{{\mathbb P}}

\newcommand{\cE}{{\mathcal E}}

\newcommand{\cL}{{\mathcal L}}
\newcommand{\cH}{{\mathcal H}}

\newcommand{\cM}{{\mathcal M}}

\newcommand{\cC}{{\mathcal C}}

\newcommand{\SU}{\mathrm{SU}}

\newcommand{\U}{\mathrm{U}}

\newcommand{\vJ}{\vec{J}}
\newcommand{\vV}{\vec{V}}

\newcommand{\id}{\mathbb{I}}

\newcommand{\be}{\begin{equation}}
\newcommand{\ee}{\end{equation}}
\newcommand{\beq}{\begin{eqnarray}}
\newcommand{\eeq}{\end{eqnarray}}
\newcommand{\bea}{\begin{eqnarray}}
\newcommand{\eea}{\end{eqnarray}}
\newcommand{\nn}{\nonumber}

\newcommand{\su}{{\mathfrak su}}

\renewcommand{\u}{{\mathfrak u}}
\newcommand{\bra}{\langle}
\newcommand{\ket}{\rangle}
\newcommand{\la}{\langle}
\newcommand{\ra}{\rangle}

\newcommand{\tr}{{\mathrm Tr}}
\newcommand{\f}{\frac}

\newcommand{\vphi}{\varphi}
\def\eps{\epsilon}

\newcommand{\hh}{{\mathbf h}}

\newcommand{\bz}{\overline{z}}

\newcommand{\Ref}[1]{(\ref{#1})}

\def\nn{\nonumber}
\def\pp{\partial}

\def\arr{\rightarrow}

\def\tphi{{\widetilde{\phi}}}

\def\vphi{\varphi}

\def\Ea{E^{(\alpha)}}
\def\Eb{E^{(\beta)}}
\def\Ua{U^{\alpha}}
\def\Ub{U^{\beta}}
\def\za{z^{(\alpha)}}
\def\zb{z^{(\beta)}}
\def\Fa{F^{(\alpha)}}
\def\Fb{F^{(\beta)}}
\def\dag{^\dagger}

\def\bz{\bar{z}}
\def\bQ{\bar{Q}}
\def\inv{{\textrm{Inv}}}
\def\cHNJ{\cH_N^{(J)}}
\def\cHN{\cH_N}
\def\tU{{{}^t U}}
\def\vcC{\vec{{\cal C}}}
\def\cHQJ{\cH^{(Q)}_J}

\def\wM{\widehat{M}}
\def\wQ{\widehat{Q}}
\def\wbQ{\widehat{\bar{Q}}}
\def\wcC{\widehat{\cC}}

\newcommand{\hcM}{{\widehat{\mathcal M}}}

\newcommand{\mat}[2]{\left(\begin{array}{#1}#2
\end{array}\right)}

\begin{document}

\title{$\U(N)$ tools for Loop Quantum Gravity: The Return of the Spinor}

\author{{\bf Enrique F. Borja}}
\affiliation{Institute for Theoretical Physics III, University of
Erlangen-N\"{u}rnberg, Staudtstra{\ss}e 7, D-91058 Erlangen (Germany)}
\affiliation{Departamento de F\'{\i}sica Te\'{o}rica and IFIC, Centro Mixto
Universidad de Valencia-CSIC. Facultad de F\'{\i}sica, Universidad de
Valencia, Burjassot-46100, Valencia (Spain)}
\author{{\bf Laurent Freidel}}
\affiliation{Perimeter Institute for Theoretical Physics, 31 Caroline St N, Waterloo ON, Canada N2L 2Y5}
\author{{\bf I\~naki Garay}}
\affiliation{Institute for Theoretical Physics III, University of
Erlangen-N\"{u}rnberg, Staudtstra{\ss}e 7, D-91058 Erlangen (Germany)}
\author{{\bf Etera R. Livine}\email{etera.livine@ens-lyon.fr}}
\affiliation{Laboratoire de Physique, ENS Lyon, CNRS-UMR 5672, 46
All\'ee d'Italie, Lyon 69007, France}

\date{\today}

\begin{abstract}

We explore the  classical setting for the $\U(N)$ framework for
$\SU(2)$ intertwiners for loop quantum gravity (LQG) and describe
the corresponding phase space in terms of spinors with the
appropriate constraints. We show how its quantization leads back to
the standard Hilbert space of intertwiner states defined as
holomorphic functionals. We then explain how to glue these
intertwiners states in order to construct spin network states as
wave-functions on the spinor phase space.
In particular, we translate the usual loop gravity holonomy observables to our classical framework. Finally, we propose how to
derive our phase space structure from an action principle which induces non-trivial dynamics for the spin network states. We conclude by applying explicitly our framework to states living on the simple 2-vertex graph and discuss the properties of the resulting Hamiltonian.


\end{abstract}

\maketitle


\tableofcontents

\section*{Introduction}

Loop quantum gravity (LQG) is now a well-established approach to
quantum gravity. It proposes a canonical framework with quantum
states -the spin network states- defining the 3d space geometry and
whose evolution in time generates space-time. The main challenges
still faced by the theory are, on the one hand, getting a full
understanding of the geometric meaning of the spin network states
both at the Planck scale and in a semi-classical regime and, on the
other hand, constructing a consistent dynamics which would lead back
to the standard dynamics of the gravitational field at large scales.
These two issues are obviously related and can not truly be solved
independently.

The present work builds on the previously introduced $\U(N)$
framework for intertwiners in loop quantum gravity
\cite{un1,un2,un3,un4}. Intertwiners are the building blocks of the
spin network states, which are constructed from gluing intertwiners
together along particular graphs. This framework was shown to be
particularly efficient in building coherent semi-classical
intertwiner states \cite{un3,un4}, which could be a useful basis to
define full coherent semi-classical spin network states. A
side-product of this approach is the possibility of reformulating
the whole LQG spin network framework in terms of spinors \cite{un3}.
From the point of view of the $\U(N)$ techniques, this comes from
the harmonic oscillators that are used to build all the operators
and Hilbert spaces and which can be understood as the quantization
of spinors. From the point of view of loop quantum gravity,
re-writing everything in terms of spinors might look like a return
to the origin since the theory was first developed in spinorial
notations. We nevertheless believe that this spinorial reformulation
is relevant to understand better the geometric meaning of the spin
network states and should be useful in studying their semi-classical
behavior and writing the quantum gravity dynamics.

This perspective is consistent with the recent ``twisted geometry"
framework developed by one of the authors and collaborators
\cite{twisted,twistor}. They explain how the classical phase space
of loop quantum gravity on a fixed graph can be expressed in terms
of spinors and show how this can be used to explore the relation
between spin network states and discrete (Regge) geometries. This is
particularly relevant to understanding the physical meaning of
spinfoam models defining the dynamics for spin networks.

In the present paper, we show how to fully recast the $\U(N)$
framework for $\SU(2)$ intertwiners in terms of spinors. More
precisely, we define the corresponding classical spinor phase space
and introduce a classical action principle from which we derive that
phase space structure. Furthermore we show how its quantization
leads to the Hilbert space of intertwiner states. These intertwiners
are built as some particular holomorphic functionals of the spinors.
We then move on to full spin network states. We explain how to glue
intertwiners together to build spin networks. This leads us to
define the classical spinor phase space behind the Hilbert space of
spin network states built on a fixed graph. In particular, we
explain how to translate the usual LQG holonomy observables in our
framework.
Then, similarly to the case of a single intertwiner, we describe the
corresponding classical action principle and discuss the possible
interaction terms we can add to the action in order to define a
non-trivial dynamics for the spin network states of quantum
geometry. Finally, we apply these techniques to spin networks on the
2-vertex graph and compare the resulting classical action principle
to the 2-vertex quantum gravity model previously constructed by some
of the authors \cite{2vertex}.

\section*{Spinors and Notations}

In this preliminary section, we introduce spinors and the related useful notations, following the previous works \cite{un3,un4,twistor}.
Considering a spinor $z$,
$$
|z\ra=\mat{c}{z^0\\z^1}, \qquad
\la z|=\mat{cc}{\bar{z}^0 &\bar{z}^1},
$$
we associate to it a geometrical 3-vector $\vec{V}(z)$, defined from the projection of the $2\times 2$ matrix $|z\ra\la z|$ onto Pauli matrices $\sigma_a$ (taken Hermitian and normalized so that $(\sigma_a)^2=\id$):
\be \label{vecV}
|z\ra \la z| = \f12 \left( {\la z|z\ra}\id  + \vec{V}(z)\cdot\vec{\sigma}\right).
\ee
The norm of this vector is obviously $|\vec{V}(z)| = \la z|z\ra= |z^0|^2+|z^1|^2$  and its components are given explicitly as:
\be
V^z=|z^0|^2-|z^1|^2,\qquad
V^x=2\,\Re\,(\bar{z}^0z^1),\qquad
V^y=2\,\Im\,(\bar{z}^0z^1).
\ee
The spinor $z$ is entirely determined by the corresponding 3-vector $\vec{V}(z)$ up to a global phase. We can give the reverse map:
\be
z^0=e^{i\phi}\,\sqrt{\f{|\vec{V}|+V^z}{2}},\quad
z^1=e^{i(\phi-\theta)}\,\sqrt{\f{|\vec{V}|-V^z}{2}},\quad
\tan\theta=\f{V^y}{V^x},
\ee
where $e^{i\phi}$ is an arbitrary phase.

Following \cite{un3}, we also introduce the map duality $\varsigma$ acting on spinors:
\be
\varsigma\begin{pmatrix}z^0\\ z^1\end{pmatrix}
\,=\,
\begin{pmatrix}-\bar{z}^1\\\bar{z}^0 \end{pmatrix},
\qquad \varsigma^{2}=-1.
\ee
This is an anti-unitary map, $\la \varsigma z| \varsigma w\ra= \la w| z\ra=\overline{\la z| w\ra}$, and we will write the related state as
$$
|z]\equiv \varsigma  | z\ra,\qquad
[z| w]\,=\,\overline{\la z| w\ra}.
$$
This map $\varsigma$ maps the 3-vector $\vec{V}(z)$ onto its opposite:
\be
|z][  z| = \f12 \left({\la z|z\ra}\id - \vec{V}(z)\cdot\vec{\sigma}\right).
\ee

Finally considering the setting necessary to describe intertwiners with $N$ legs, we consider $N$ spinors $z_i$ and their corresponding 3-vectors $\vV(z_i)$.
Typically, we can require that the $N$ spinors satisfy a closure
condition, i.e  that the sum of the corresponding 3-vectors
vanishes, $\sum_i \vec{V}(z_i)=0$. Coming back to the definition of
the 3-vectors $\vV(z_i)$, the closure condition is easily translated
in terms of $2\times 2$ matrices:
\be
\sum_i |z_i\ra \la z_i|=A(z)\id,
\qquad\textrm{with}\quad
A(z)\equiv\f12\sum_i \la z_i|z_i\ra=\f12\sum_i|\vec{V}(z_i)|.
\ee
This further translates into quadratic constraints on the spinors:
\be
\sum_i z^0_i\,\bar{z}^1_i=0,\quad
\sum_i \left|z^0_i\right|^2=\sum_i \left|z^1_i\right|^2=A(z).
\label{closure}
\ee
In simple terms, it means that the two components of the spinors, $z^0_i$ and $z^1_i$, are orthogonal $N$-vectors of equal norm.

\section{Overview of the $\U(N)$ Structure of Interwiners}

Here, we quickly review the $\U(N)$ formalism for $\SU(2)$
intertwiners in loop quantum gravity. This framework was introduced
and improved in a series of papers \cite{un1,un2,un3,un4,2vertex}.
More precisely, intertwiners with $N$ legs are $\SU(2)$-invariant
states in the tensor product of $N$ (irreducible) representations of
$\SU(2)$. Then the  basic tool used to define the $\U(N)$ formalism
is the Schwinger representation of the $\su(2)$ Lie algebra in terms
of a pair of harmonic oscillators. Since we would like to describe
the tensor product of $N$ $\SU(2)$-representations, we will need $N$
copies of the $\su(2)$-algebra  and thus we consider $N$ pairs of
harmonic oscillators $a_i,b_i$ with $i$ running from 1 to $N$.

The local $\su(2)$ generators acting on each leg $i$ are defined as quadratic operators:
\be
J^z_i=\f12(a\dag_i a_i-b\dag_ib_i),\qquad
J^+_i=a\dag_i b_i,\qquad
J^-_i=a_i b\dag_i,\qquad
E_i=(a\dag_i a_i+b\dag_ib_i).
\ee
The $J_i$'s satisfy the standard commutation algebra while the total
energy $E_i$  is a Casimir operator:
\be
[J^z_i,J^\pm_i]=\pm J^\pm_i,\qquad
[J^+_i,J^-_i]=2J^z_i,\qquad
[E_i,\vec{J}_i]=0.
\ee
The operator $E_i$ is the total energy carried by the pair of
oscillators $a_i,b_i$ and simply gives twice the spin $2j_i$ of the
corresponding $\SU(2)$-representation. Indeed, we can easily express
the standard $\SU(2)$ Casimir operator in terms of this energy:
\be
\vJ_i^2\,=\, \f{E_i}2\left(\f{E_i}2+1\right)\,=\, \f{E_i}4\left({E_i}+2\right).
\ee
In the context of loop quantum gravity, the spin $j_i$ given as the value of the operator $E_i/2$ defines the area associated to the leg $i$ of the intertwiner.

Then we look for operators invariant under global $\SU(2)$ transformations generated by
$\vJ\,\equiv\,\sum_i \vJ_i$. The key result, which is the starting point of the $\U(N)$ formalism, is that we can identify quadratic invariant operators acting on pairs of (possibly equal) legs $i,j$ \cite{un1,un3}:
\be
\label{defE}
E_{ij}=a\dag_ia_j+b\dag_ib_j, \qquad
E_{ij}\dag=E_{ji},
\ee
\be
\label{defF}
F_{ij}=(a_i b_j - a_j b_i),\qquad
F_{ji}=-F_{ij}.
\ee
These operators $E,F,F\dag$ form a closed algebra:
\bea
\label{commEF}
{[}E_{ij},E_{kl}]&=&
\delta_{jk}E_{il}-\delta_{il}E_{kj},\\
{[}E_{ij},F_{kl}] &=& \delta_{il}F_{jk}-\delta_{ik}F_{jl},\qquad
{[}E_{ij},F_{kl}^{\dagger}] = \delta_{jk}F_{il}^{\dagger}-\delta_{jl}F_{ik}^{\dagger}, \nn\\
{[} F_{ij},F^{\dagger}_{kl}] &=& \delta_{ik}E_{lj}-\delta_{il}E_{kj} -\delta_{jk}E_{li}+\delta_{jl}E_{ki}
+2(\delta_{ik}\delta_{jl}-\delta_{il}\delta_{jk}), \nn\\
{[} F_{ij},F_{kl}] &=& 0,\qquad {[} F_{ij}^{\dagger},F_{kl}^{\dagger}] =0.\nn
\eea
First focusing on the $E_{ij}$ operators, their commutators close
and they form a $\u(N)$-algebra. This is why this formalism has been
dubbed the $\U(N)$ framework for loop quantum gravity
\cite{un1,un2}. The diagonal operators are equal to the previous
operators giving the energy on each leg, $E_{ii}=E_i$. Then the
value of the  total energy $E\,\equiv \sum_i E_i$ gives twice the
sum of all spins $2\times\sum_i j_i$, i.e twice the total area in
the context of loop quantum gravity.

The $E_{ij}$-operators change the energy/area carried by each leg, while still conserving the total energy, while the operators $F_{ij}$ (resp. $F\dag_{ij}$) will decrease (resp. increase) the total area $E$ by 2:
\be
[E,E_{ij}]=0,\qquad
[E,F_{ij}]=-2F_{ij},\quad
[E,F\dag_{ij}]=+2F\dag_{ij}.
\ee
This suggests to decompose the Hilbert space of $N$-valent intertwiners into subspaces of constant area:
\be
\cH_N=\bigoplus_{\{j_i\}} \inv\left[\otimes_{i=1}^NV^{j_i}\right]
=\bigoplus_{J\in\N}\bigoplus_{\sum_ij_i=J} \inv\left[\otimes_{i=1}^NV^{j_i}\right]
=\bigoplus_J \cH_N^{(J)},
\ee
where $V^{j_i}$ denote the Hilbert space of the irreducible $\SU(2)$-representation of spin $j_i$, spanned by the states  of the oscillators $a_i,b_i$ with fixed total energy $E_i=2j_i$.

It was proven in \cite{un2} that each subspace $\cHNJ$ of $N$-valent intertwiners with fixed total area $J$ carries an irreducible representation of $\U(N)$ generated by the $E_{ij}$ operators. These are representations with Young tableaux given by two horizontal lines with equal numbers of cases ($J$). More constructively, we can characterize them by their highest weight vector $v_N^{(J)}$:
\be
E_1\,|v_N^{(J)}\ra\,=\,J\,|v_N^{(J)}\ra,\quad
E_2\,|v_N^{(J)}\ra\,=\,J\,|v_N^{(J)}\ra,\quad
E_{k\ge 3}\,|v_N^{(J)}\ra\,=\,0,\quad
E\,|v_N^{(J)}\ra\,=\,2J\,|v_N^{(J)}\ra,\qquad
E_{i<j}\,|v_N^{(J)}\ra\,=\,0\,.
\ee
These highest weight vectors define bivalent intertwiners where all the area is carried by two legs $i=1,2$ while all the other legs carried the trivial $\SU(2)$-representation $j_{k\ge 3}=0$.
Then the operators $E_{ij}$ allow to navigate from state to state within each subspace $\cHNJ$. On the other hand, the operators $F_{ij},\,F\dag_{ij}$ allow to go from one subspace $\cHNJ$ to the next $\cHN^{(J\pm 1)}$, thus endowing the full space of $N$-valent intertwiners with a Fock space structure with creation operators $F\dag_{ij}$ and annihilation operators $F_{ij}$.

Finally, the identification of the highest vectors was made possible by realizing that the operators $E_{ij}$ satisfy quadratic constraints, which can then be turned by conditions relating the quadratic $\U(N)$-Casimir operator to the total area $E$ \cite{un2}. Then it was realized that the whole set of operators $E_{ij},F_{ij},F\dag_{ij}$ satisfy quadratic constraints \cite{2vertex}:
\be
\forall i,j,\quad
\sum_k E_{ik}E_{kj}=E_{ij} \left(\f E2+N-2\right),
\label{constraint0}
\ee
\beq
&&\sum_k F^\dagger_{ik}E_{jk} = F^\dagger_{ij}\, \frac{E}{2}, \qquad\qquad\quad
\sum_k E_{jk} F^\dagger_{ik} = F^\dagger_{ij}\left(\frac{E}{2}+N-1\right),\label{constraint1}\\
&&\sum_k E_{kj}F_{ik} = F_{ij}\, \left(\frac{E}{2}-1\right), \qquad
\sum_k F_{ik} E_{kj}  = F_{ij}\left(\frac{E}{2}+N-2\right),\label{constraint2}\\
&&\sum_k F^\dagger_{ik}F_{kj} = E_{ij}
\left(\frac{E}{2}+1\right),\qquad
\sum_k F_{kj}F\dag_{ik} = (E_{ij}+2\delta_{ij})
\left(\frac{E}{2}+N-1\right)\,.\label{constraint3}
\eeq
As already noticed in \cite{2vertex}, these relations look a lot like constraints on the multiplication of two matrices $E_{ij}$ and $F_{ij}$. This is the point which we will explore further in the present paper, and we will show that the operators $E_{ij}$ and $F_{ij}$ are truly the quantization of the matrix elements of two $N\times N$ classical matrices built from a set of $2N$ spinors. This will allow to explore further the representation of the intertwiner space $\cHN$ as the $L^2$ over the Grassmanian space $U(N)/(\U(2)\times\U(N-2))$ introduced in \cite{un2,un3}.

\section{The Classical Setting for Intertwiners}

\subsection{The Matrix Algebra}

Drawing inspiration  from the operators $E_{ij}$ and $F_{ij}$ and the quadratic constraints relating them, our goal is to describe the classical system behind the Hilbert space of $\SU(2)$-intertwiners. Thus we consider two $N\times N$ matrices, a Hermitian matrix and an antisymmetric one:
\be
M=M\dag,\quad {}^tQ=-Q.
\ee
We assume that they satisfy the same constraints
(\ref{constraint0}-\ref{constraint3}) that the operators $E_{ij}$
and $F_{ij}$, up to terms which we identify as coming from quantum
ordering :
\beq
&&M^2=\f{\tr M}2\,M,\qquad
Q\bar{Q}=-\f{\tr M}2\,M,\nn\\
&&MQ=\f{\tr M}2\,Q,\qquad \bar{Q}M=\f{\tr M}2\,\bar{Q},
\eeq
where $Q$ actually plays the role of $F\dag$ while $\bar{Q}$ corresponds to $F$.

Let us now solve these equations and parameterize the space of solutions to these constraints.

\begin{res}
The quadratic constraints on $M$ and $Q$, together with the requirement of Hermiticity of $M$ and anti-symmetry of $Q$, entirely fix these two matrices up to a global $\U(N)$ transformation and a relative phase:
\beq
&&M=\lambda\,U\Delta U^{-1},\qquad\,\, \Delta=\mat{cc|c}{1 & & \\ &1 & \\ \hline && 0_{N-2}}\,, \\
&&Q=e^{i\theta}\lambda\,U\Delta_\eps \tU,\qquad \Delta_\eps=\mat{cc|c}{ & 1& \\ -1& & \\ \hline & &0_{N-2}}\,, \nn
\eeq
where $U$ is a unitary matrix $U\dag U=\id$, $\lambda\in\R$ and $\exp(i\theta)$ is an arbitrary phase.
\end{res}

\begin{proof}

Let us start with the trivial case when $\tr M=0$. Then it is obvious to see that $M=Q=0$. Let us thus assume that $\tr M\ne 0$ and let us define $\lambda=(\tr M)/2$.

The equation $M^2=\lambda M$ implies that the matrix $M$ is a
projector, with two eigenvalues 0 and $\lambda$. Then using that
$\lambda=(\tr M)/2$, we can conclude that its rank is two. Thus we
can write $M=\lambda\,U\Delta U^{-1}$, in terms of a unitary matrix
$U\in\U(N)$ defining an orthonormal basis diagonalizing $M$. More
precisely, calling $|e_i\ra$ the canonical basis for $N$-vectors,
the basis $U\,|e_i\ra$ diagonalizes $M$:
$$
M\,U|e_1\ra=\lambda\,U|e_1\ra,\quad
M\,U|e_2\ra=\lambda\,U|e_2\ra,\quad
M\,U|e_{k\ge 3}\ra=\,0.
$$
The next step is to determine the matrix $Q$ in terms of $\lambda$
and $U$. We first apply the condition that
$\bar{Q}M=\lambda\,\bar{Q}$, this implies that:
$$
\bar{Q}\,U|e_{k\ge 3}\ra=\,0,
$$
which is equivalent to $Q\,\bar{U}|e_{k\ge 3}\ra=\,0$. Then we can use the condition $MQ=\lambda Q$ to determine the action of $Q$ on the first two basis vectors:
$$
\forall i=1,2,\quad M\,Q\,\bar{U}|e_i\ra=\lambda\,Q\,\bar{U}|e_i\ra.
$$
Looking at the state $\bar{U}|e_1\ra$, this means that either $Q\,\bar{U}|e_1\ra=0$ or that $Q\,\bar{U}|e_1\ra$ is in the subspace generated by $U|e_1\ra$ and $U|e_2\ra$. The first possibility is impossible, since it would imply that $MU|e_1\ra=0$ due to the condition $Q\bar{Q}=-\lambda M$. Thus we write:
$$
Q\,\bar{U}|e_1\ra=\alpha U|e_1\ra+\beta U|e_2\ra.
$$
Moreover, since $Q$ is antisymmetric, we have $\la Ue_1|Q|\bar{U}e_1\ra=0$ and thus the coefficient $\alpha$ vanishes. Further using the antisymmetry property, we have $\la Ue_1|Q|\bar{U}e_2\ra=-\la Ue_2|Q|\bar{U}e_1\ra$, thus we get:
$$
Q\,\bar{U}|e_1\ra=\beta U|e_2\ra,\qquad
Q\,\bar{U}|e_2\ra=-\beta U|e_1\ra.
$$
Finally, using the last condition $Q\bar{Q}=-\lambda M$, we can compute the value of the coefficient $\beta$:
$$
|\beta|^2\,U|e_1\ra=-Q\bar{Q}\,U|e_1\ra=\lambda M\,U|e_1\ra=\lambda^2\,U|e_1\ra
\qquad\Rightarrow\quad
\beta=e^{i\theta}\lambda,
$$
where $\theta$ is an arbitrary angle. This allows to conclude that: $Q=e^{i\theta}\lambda\,U\Delta_\eps \tU$ since $\tU=\bar{U}^{-1}$.

Reversely, it is easy to check that the resulting expressions for
$M$ and $Q$ in terms of $U,\lambda,\theta$ always satisfy the
quadratic constraints which we started from.

\end{proof}

From now on, using the $U(1)$ freedom of chosing $U$, we will set the phase $\theta$ to 0 and define the two matrices as:
\be
M=\lambda\,U\Delta U^{-1},\qquad Q=\lambda\,U\Delta_\eps \tU.
\ee
Comparing with the $\U(N)$ framework for intertwiners reviewed in
the previous section, the rank-2 matrix $\Delta$ plays the role of
the highest weight vector, from which we will get the full space of
states by acting on it with $\U(N)$ transformations. Looking at the
stabilizer group for the diagonal matrix $\Delta$, we see that $M$
is invariant under:
\be
U\arr UV,\quad \forall V\in\U(2)\times\U(N-2),
\ee
and therefore the matrix $M$ exactly parameterizes the coset space $\U(N)/\U(2)\times\U(N-2)$, which was already identified in \cite{un2,un3} as the classical space behind $N$-valent $\SU(2)$ intertwiners.
Similarly looking at the stabilizer group for $\Delta_\eps$, we realize that $Q$ is invariant under a slightly smaller subgroup:
\be
U\arr UV,\quad \forall V\in\SU(2)\times\U(N-2).
\ee
Therefore, the space of functions $f(Q)$ invariant under multiplication by a phase, $f(Q)=f(e^{i\theta}Q)$, is isomorphic to the space of functions on the Grassmannian space $\U(N)/\U(2)\times\U(N-2)$. This is consistent with the fact that the quadratic conditions on $M$ and $Q$ only determine the matrix $Q$ up to a phase.

Finally we will also require the positivity of the matrix $M$, i.e $\lambda\ge 0$. This reflects the positivity of the energy/area $E$ at the quantum level. So that we now work with $\lambda\in\R^+$.

\subsection{From $\U(N)$ Matrices to Spinors and the Closure Condition}

We start by writing explicitly the two matrices $M$ and $Q$ in terms
of the matrix elements of the unitary transformation $U$:
\be
M_{ij}=\lambda(u_{i1}\bar{u}_{j1}+u_{i2}\bar{u}_{j2}),
\qquad
Q_{ij}=\lambda(u_{i1}u_{j2}-u_{i2}u_{j1}).
\ee
These expressions only involve the first columns of the matrix $U$. This comes from the definition of the diagonal matrices $\Delta$ and $\Delta_\eps$, and the resulting invariance under the $\U(N-2)$ subgroup.
Comparing these equations with the definitions
(\ref{defE}-\ref{defF}) of the operators $E_{ij}$ and $F\dag_{ij}$,
we see that the matrix element $u_{i1}$ corresponds to the operator
$a\dag_i$. Following this logic of a classical-quantum
correspondence, we define the spinors $z_i$ as the rescaled first
two columns of the $U$-matrix:
\be
z_i\,\equiv\,\mat{c}{\bar{u}_{i1}\sqrt{\lambda}\\ \bar{u}_{i2}\sqrt{\lambda}}.
\ee
The matrices $M$ and $Q$ are easily expressed in terms of these
spinors:
\be
M_{ij}=\la z_i |z_j \ra=\overline{\la z_j |z_i \ra},
\qquad
Q_{ij}=\la z_j |z_i]=\overline{[ z_i |z_j \ra}=-\overline{[ z_j |z_i \ra}.
\ee
The matrix elements $Q_{ij}$ are clearly anti-holomorphic in the
$z_i$'s while the $M_{ij}$'s mix both holomorphic and
anti-holomorphic components.
With $M$ and $Q$ written as such, the quadratic constraints on $M$ and $Q$ are exactly the relations between the matrices $\la z_i |z_j \ra$ and $[ z_j |z_i \ra$ written in \cite{un3}.
The spinors $z_i$ are not entirely free, since they come from the
unitary matrix $U$. The only constraint is that the two vectors
$u_{i1}$ and $u_{i2}$ are part of an orthonormal matrix, that is
that they are of unit-norm and orthogonal:
\be
U\dag U =\id \qquad\Rightarrow\quad
\sum_i |u_{i1}|^2=\sum_i |u_{i2}|^2=1,\quad
\sum_i u_{i1}\bar{u}_{i2}=0.
\ee
This is easy translated into conditions on the spinors:
\be
\sum_i |z^0_i|^2=\sum_i |z^1_i|^2=\lambda,\quad
\sum_i z^0_i \bz^1_i=0.
\ee
Checking out the short preliminary section about spinors, we see that these conditions correspond exactly to the {\it closure constraints} on the spinors $z_i$, thus they are equivalent to the following conditions:
\be
\sum_i \vV(z_i)=0,\quad
\sum_i |z_i\ra\la z_i|=\lambda\id,
\qquad
\f12\sum_i |\vV(z_i)| =\f12\sum_i \la z_i| z_i\ra=\lambda.
\ee
Thus the requirement of unitarity, that our matrix $U$ lays in $\U(N)$, is equivalent to the closure conditions on our $N$ spinors.
We could relax these closure conditions by dropping the requirement of unitarity, but this would break the quadratic constraints that $M$ and $Q$ satisfy.

Let us introduce the matrix elements of the $2\times 2$ matrix $\sum_i |z_i\ra\la z_i|$~:
\be
\cC_{ab}=\sum_i z^a_i \bz^b_i.
\ee
Then the unitary or closure conditions are written very simply:
\be
\cC_{00}-\cC_{11}=0,\quad \cC_{01}=\cC_{10}=0.
\ee

\subsection{Phase Space and $\SU(2)$ Invariance}

Let us introduce a simple Poisson bracket on our space of $N$ spinors:
\be
\label{bracketz}
\{z^a_i,\bz^b_j\}\,\equiv\,i\,\delta^{ab}\delta_{ij},
\ee
with all other brackets vanishing, $\{z^a_i,z^b_j\}=\{\bz^a_i,\bz^b_j\}=0$. This is exactly the Poisson bracket for $2N$ decoupled harmonic oscillators.

We start by checking that the closure conditions generates global $\SU(2)$ transformations on the $N$ spinors. First, we can compute the Poisson brackets between the various components of the $\cC$-constraints~:
\beq
\label{commC}
&&\{\cC_{00}-\cC_{11},\cC_{01}\}=-2i\cC_{01},\quad
\{\cC_{00}-\cC_{11},\cC_{10}\}=+2i\cC_{10},\quad
\{\cC_{10},\cC_{01}\}=i(\cC_{00}-\cC_{11}),\\
&&\{\tr \cC,\cC_{00}-\cC_{11}\}=\{\tr \cC,\cC_{01}\}=\{\tr \cC,\cC_{10}\}=0.\nn
\eeq
These four components $\cC_{ab}$ do indeed form a closed $\u(2)$ algebra with the three closure conditions $\cC_{00}-\cC_{11}$, $\cC_{01}$ and $\cC_{10}$ forming the $\su(2)$ subalgebra.
Thus we will write $\vec{\cC}$ for these three $\su(2)$-generators with $\cC^z\equiv\cC_{00}-\cC_{11}$ and $\cC^+=\cC_{10}$ and $\cC^-=\cC_{01}$.
We can further check their commutator with the spinors themselves:
\be
\begin{array}{lll}
\{\cC_{00}-\cC_{11},z^0_i\}= -i\,z^0_i,\quad & \{\cC_{01},z^0_i\}=0,
& \{\cC_{10},z^0_i\}=-iz^1_i,
\\
\{\cC_{00}-\cC_{11},z^1_i\}= +i\,z^1_i, &
\{\cC_{01},z^1_i\}=-iz^0_i,\quad & \{\cC_{10},z^1_i\}=0,
\end{array}
\ee
which indeed generates the standard $\SU(2)$ transformations on the $N$ spinors.
The three closure conditions $\vec{\cC}$ will actually  become the generators $\vJ$ at the quantum level, while the operator $\tr\,\cC$ will correspond to the total energy/area $E$.

\medskip

We also compute the Poisson brackets of the $M_{ij}$ and $Q_{ij}$ matrix elements:
\beq
&&\{M_{ij},M_{kl}\}=i(\delta_{kj}M_{il}-\delta_{il}M_{kj}), \label{commM}\\
&&\{M_{ij},Q_{kl}\}=i(\delta_{jk}Q_{il}-\delta_{jl}Q_{ik}),\nn\\
&&\{Q_{ij},Q_{kl}\}=0,\nn\\
&&\{\bar{Q}_{ij},Q_{kl}\}=i(\delta_{ik}M_{lj}+\delta_{jl}M_{ki}-\delta_{jk}M_{li}-\delta_{il}M_{kj}),\nn
\eeq
which reproduces the expected commutators \Ref{commEF} up to the $i$-factor. We further check that these variables commute with the closure constraints generating the $\SU(2)$ transformations:
\be
\{\vcC,M_{ij}\}=\{\vcC,Q_{ij}\}=0.
\ee
Finally, we look at their commutator with $\tr \,\cC$. On the one hand, we have:
\be
\{\tr\,\cC,M_{ij}\}=0,
\ee
which confirms that the matrix $M$ is invariant under the full $\U(2)$ subgroup.
On the other hand, we compute:
\be
\{\tr\,\cC,Q_{ij}\}=\{\sum_k M_{kk},Q_{ij}\}=\,+2i\,Q_{ij},
\ee
which means that $\tr\,\cC$ acts as a dilatation operator on the $Q$ variables, or reversely that the $Q_{ij}$ acts as creation operators for the total energy/area variable $\tr\,\cC$.

\subsection{Action and Matrix model}

We can derive the previous Poisson bracket from an action principle,
which  directly defines the classical phase space associated to
$\SU(2)$ intertwiners. In terms of the spinor variables, the action
simply reads:
\beq
S_0[z_i]
&\equiv&
\int dt \,\left(
\sum_{i,a} iz^a_i\pp_t\bz^a_i -\Lambda^{ab}\cC_{ab}
\right)
\,=\,
\int dt \,\left(
\sum_{i,a} iz^a_i\pp_t\bz^a_i -\Lambda^{ab}\sum_i z_i^a\bz_i^b
\right), \\
&=&
\int dt \,\left(
\sum_{i} -i\la z_i|\pp_tz_i \ra+ \la z_i|\Lambda| z_i \ra
\right),\nn
\eeq
where the $2\times 2$ matrix elements $\Lambda_{ab}$ are Lagragian multipliers satisfying $\tr\,\Lambda=\sum_a\Lambda_{aa}=0$ and enforcing the closure constraints $\vcC=0$. As we have seen in the previous sections, the three constraints $\vcC$ are first class constraints, they generate global $\SU(2)$ transformations on the spinors. We must both impose and solve these closure constraints and identify $\SU(2)$-invariant observables on the space of constrained spinors.

This is the free action defining only the classical kinematics on
the intertwiner space described in terms of spinors. We can define
dynamics by adding an interaction term to the action:
$$
S[z_i] \,\equiv\, S_0[z_i]+I[z_i] \,=\, \int dt \,\left( \sum_{i,a}
iz^a_i\pp_t\bz^a_i -\Theta^{ab}\cC_{ab} -H[z_i] \right)\,,
$$
where $H[z_i]$ would be the Hamiltonian of the system defining how intertwiners evolve.

\medskip

We can re-write this action principle in terms of the initial
unitary matrix $U$ and the parameter $\lambda\in\R^+$:
\beq
S_0[U,\lambda] &\equiv& \int dt \,\left[ -i\,\tr\,
\sqrt{\lambda}U\Delta \pp_t (\sqrt{\lambda}U\dag) \,-\,\tr
\Theta\,(UU\dag-\id)
\right] \\
&=& \int dt \,\left( -i\,\lambda \tr\,U\Delta \pp_t U\dag \,
-i\,\pp_t\lambda \,-\,\tr \Theta\,(UU\dag-\id) \right), \nn
\eeq
where the $N\times N$ matrix $\Theta$ is a Lagrange multiplier enforcing that the matrix $U$ is unitary.

A first remark is that the kinematical term
$\sqrt{\lambda}\pp_t\sqrt{\lambda}$ is a total derivative and does
not induce any evolution, thus the dynamics of the variable
$\lambda$ is entirely determined by its coupling to $U$ through the
kinematical term $\lambda \tr\,U\Delta \pp_t U\dag$. Therefore, if
the unitary matrix $U$ does not evolve, then $\lambda$ is frozen
too.
Then, we see that the action is invariant under the left $\U(N)$ action:
$$
U\arr VU, \quad V\in\U(N),
$$
for a constant unitary matrix $V$ (independent from $t$), but is only invariant under the right action of the stabilizer subgroup $\U(2)\times\U(N-2)$:
$$
U\arr UV, \quad V\in\U(2)\times\U(N-2).
$$

If we want to add dynamics to this system, it is natural to require that the interaction term be also invariant under the same symmetries. This greatly constrains the possible terms of a Hamiltonian. Indeed, we are left with polynomials of $\tr (\lambda\,U\Delta U^{-1})^k \,=\,
\tr\, M^k=2\lambda^k$, which are simply polynomials in $\lambda$. Since the variable $\lambda$ is not really dynamical, we conclude that there are no truly non-trivial invariant dynamics for a single intertwiner.

We could bypass this conclusion by allowing Hamiltonian operators that break the $\U(N)$ symmetry. We do not really see the purpose of such procedure, although it could be used to model the coupling of a single intertwiner to an external source breaking the $\U(N)$ invariance. On the other hand, we will see in section.\ref{dynamics} that we can have non-trivial dynamics as soon as we work with many intertwiners when considering true spin network states on an arbitrary graph.

%
%

\medskip

Finally, we end this section stressing that we have managed to reformulate the classical setting of a single $\SU(2)$ intertwiner as a unitary matrix model, which was the original goal of the paper \cite{un1} which introduces the $\U(N)$ framework for $\SU(2)$ intertwiners.

\subsection{Intertwiners as (Anti-)Holomorphic Functionals}
\label{antiholo}

%

Now that we have fully characterize the phase space associated to the spinors $z_i$ and the variables $M_{ij},Q_{ij}$, we can proceed to the quantization.

\begin{figure}[h]
\begin{center}
\includegraphics[height=30mm]{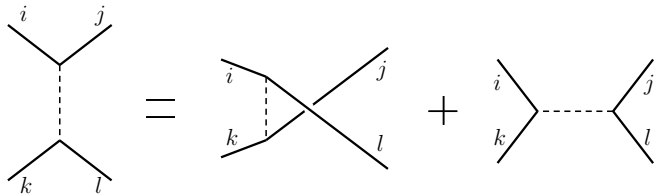}
\caption{Focusing on the four legs $(i,j,k,l)$ of the intertwiner, the Pl\"ucker relation $Q_{ij}Q_{kl}=Q_{il}Q_{kj}+Q_{ik}Q_{jl}$ on the $Q$-variables becomes the (standard) recoupling relation for $\SU(2)$ intertwiners (more precisely, for holonomy operators acting on $\SU(2)$ intertwiners). This relation is often used in Loop Quantum Gravity  when still using states defined as products of Wilson loops instead of spin network states.
\label{QQfig}}
\end{center}
\end{figure}

The most natural choice is to consider polynomials in the $Q_{ij}$
matrix elements. More precisely, we introduce the Hilbert spaces
$\cHQJ$ of homogeneous polynomials in the $Q_{ij}$ of degree $J$:
\be
\cHQJ \,\equiv\, \{P \in \pP[Q_{ij}] \,|\quad P(\rho
Q_{ij})\,=\,\rho^J\,P(Q_{ij}),\,\forall\rho\in\C  \}\,.
\ee
These are polynomials completely anti-holomorphic in the spinors $z_i$ (or holomorphic in $\bz_i$) and of order $2J$.
Let us point out that the variables $Q_{ij}$ are not independent,
since they are expressed in terms of the spinors $z_i$. Resultingly,
they are related to each other by the Pl\"ucker relations as already
noticed in \cite{un3}:
\be
Q_{ij}=\,\bz_i^0\bz_j^1-\bz_i^1\bz_j^0
\qquad\Rightarrow\quad
Q_{ij}Q_{kl}=Q_{il}Q_{kj}+Q_{ik}Q_{jl}.
\ee
Interestingly, this can be interpreted as the recoupling relation between interchanging the four legs $(i,j,k,l)$ of the intertwiner (see fig.\ref{QQfig}).

{\it }

Our claim is that these Hilbert spaces $\cHQJ$ are isomorphic to the Hilbert space $\cHNJ$ of $N$-valent intertwiners with fixed total area $J$. To this purpose, we will construct the explicit representation of the operator quantizing $M_{ij}$ and $Q_{ij}$ on the spaces $\cHQJ$ and show that they match the actions of the $\U(N)$ operators $E_{ij}$ and $F\dag_{ij}$ which we described earlier.
Our quantization relies on quantizing the $\bz_i$ as multiplication operators while promoting $z_i$ to a derivative operator:
\be
\widehat{\bz}_i^a
\,\equiv\,
{\bz}_i^a\,\times\,,\qquad
\widehat{z}_i^a
\,\equiv\,
\f{\pp}{\pp \bz_i^a},
\ee
which satisfies the commutator $[\hat{z},\hat{\bz}]=1$ as expected
for the quantization of the classical bracket $\{z,\bz\}=i$. Then,
we quantize the matrix elements $M_{ij}$ and $Q_{ij}$ and the
closure constraints following this correspondence:
\beq
\widehat{M}_{ij} &=&
{\bz}_i^0\f{\pp}{\pp \bz_j^0}+{\bz}_i^1\f{\pp}{\pp \bz_j^1}\,, \\
\widehat{Q}_{ij} &=&
\bz_i^0\bz_j^1-\bz_i^1\bz_j^0 \,=\, Q_{ij}\,, \\
\widehat{\bQ}_{ij} &=&
\f{\pp^2}{\pp \bz_i^0\pp \bz_j^1}-\f{\pp^2}{\pp \bz_i^1\pp \bz_j^0}\,, \\
\widehat{\cC}_{ab} &=&
\sum_k{\bz}_k^b\f{\pp}{\pp \bz_k^a}.
\eeq
It is straightforward to check that the $\wcC_{ab}$ and the
$\wM_{ij}$ respectively form a $\u(2)$ and a $\u(N)$ Lie algebra, as
expected:
\be
[\wcC_{ab},\wcC_{cd}]\,=\, \delta_{ad}\wcC_{cb}-\delta_{cb}\wcC_{ad},\qquad
[\wM_{ij},\wM_{kl}]\,=\, \delta_{kj}\wM_{il}-\delta_{il}\wM_{kj},\qquad
[\wcC_{ab},\wM_{ij}]\,=\,0.
\ee
which amounts to multiply the Poisson bracket \Ref{commC} and \Ref{commM}  by $-i$.
Then, we first check the action of the closure constraints on functions of the variables $Q_{ij}$~:
$$
\widehat{\vcC}\,Q_{ij}\,=\, 0 ,\qquad
\widehat{(\tr\,\cC)}\,Q_{ij}\,=\, 2Q_{ij},
$$
\be
\forall P\in\cHQJ=\pP_J[Q_{ij}],\qquad
\widehat{\vcC}\,P(Q_{ij})\,=\,0,\qquad
\widehat{(\tr\,\cC)}\,P(Q_{ij})\,=\, 2J\,P(Q_{ij}),
\ee
so that our wavefunctions $P\in\cHQJ$ are $\SU(2)$-invariant (vanish under the closure constraints) and are eigenvectors of the $\tr\,\cC$-operator with eigenvalue $2J$.

Second, we check that the operators $\wM$ and $\widehat{(\tr\,\cC)}$ satisfy the same quadratic constraints on the Hilbert space $\cHQJ$ (i.e assuming that the operators acts on $\SU(2)$-invariant functions vanishing under the closure constraints) that the $\u(N)$-generators $E_{ij}$:
\be
\widehat{(\tr\,\cC)}=\sum_k \wM_{kk},\qquad
\sum_k \wM_{ik}\wM_{kj}
\,=\,
\wM_{ij}\left(\f {\widehat{(\tr\,\cC)}}2 +N-2\right),
\ee
which allows us to get the value of the (quadratic) $U(N)$-Casimir operator on the space $\cHQJ$:
$$
\sum_{ik}\wM_{ik}\wM_{ki}\,=\,
\widehat{(\tr\,\cC)}\left(\f {\widehat{(\tr\,\cC)}}2 +N-2\right)
\,=\,
2J(J+N-2).
$$
Thus, we can safely conclude that this provides a proper quantization of our spinors and $M$-variables, which matches exactly with the  $\u(N)$-structure on the intertwiner space (with the exact same ordering):
\be
\cHQJ \sim\cHNJ,\quad
\wM_{ij}\,=\,E_{ij},\quad
\widehat{(\tr\,\cC)}\,=\,E.
\ee

Now, turning to the $\wQ_{ij}$-operators, it is straightforward to check that they have the exact same action that the $F\dag_{ij}$ operators, they satisfy the same Lie algebra commutators \Ref{commEF} and the same quadratic constraints (\ref{constraint1}-\ref{constraint3}). Clearly, the simple multiplicative action of an operator $\wQ_{ij}$ send a polynomial in $\pP_J[Q_{ij}]$ to a polynomial in $\pP_{J+1}[Q_{ij}]$. Reciprocally, the derivative action of $\wbQ_{ij}$ decreases the degree of the polynomials and maps  $\pP_{J+1}[Q_{ij}]$ onto $\pP_J[Q_{ij}]$.

Finally, let us look at the scalar product on whole space of
polynomials $\pP[Q_{ij}]$. In order to ensure the correct Hermicity
relations for $\wM_{ij}$ and $\wQ_{ij},\wbQ_{ij}$, it seems that we
have a unique\footnote{ If we ask to recover only the hermiticity
relation for $\wM_{ij}$ then we can use any function of $\sum_i \la
z_i|z_i\ra$ as a measure instead of the exponential. Asking in turns
that $\wQ$ and $\wbQ$ are hermitian  conjugate fixes entirely the
measure up to a scale.}
 measure (up to a global factor):
\be
\label{scalarp}
\forall \phi,\psi\in \pP[Q_{ij}],\quad
\la\phi|\psi\ra \,\equiv\,
\int \prod_id^4z_i\,e^{-\sum_i \la z_i|z_i\ra}\,
\overline{\phi(Q_{ij})}\,\psi(Q_{ij})\,.
\ee
Then it is easy to check that we have $\wM\dag_{ij}=\wM_{ji}$ and $\wQ\dag_{ij}=\wbQ_{ij}$ as wanted.

It is easy to see that the spaces of homogeneous polynomials
$\pP_J[Q_{ij}]$ are  orthogonal with respect to this scalar product.
The quickest way to realize that this is true is to consider the
operator $\widehat{(\tr\,\cC)}$, which is Hermitian with respect to
this scalar product and takes different values on the spaces
$\pP_J[Q_{ij}]$ depending on the value of $J$. Thus these spaces
$\pP_J[Q_{ij}]$ are orthogonal to each other\footnote{If
$\phi_{J}(Q_{ij})\,\psi_{J}(Q_{ij})$ are homogeneous of degree $J$
we can express this scalar product as an integral over the
grassmanian:
\be
\la\phi_{J}|\psi_{J}\ra \,=\, (N+J-1)!(N+J-2)! \int_{G_{2,N}} \,
{\overline{\phi_{J}(Q_{ij})}\,\psi_{J}(Q_{ij})}\,,
\ee
where\, $G_{2,N}=\{ |z_{i}\ket_{i=1\cdots n} | \sum_{i}
|z_{i}\ket\bra z_{i}| =1\}$ .}.

This concludes our quantization procedure thus showing that the
intertwiner space for $N$ legs and fixed total area $J=\sum_i j_i$
can be seen as the space of homogeneous polynomials in the $Q_{ij}$
variables  with degree $J$. This provides us with a description of
the intertwiners as wave-functions anti-holomorphic in the spinors
$z_i$ (or equivalently holomorphic in $\bz_i$) constrained by the
closure conditions.
In particular, the highest weight vector of the $\U(N)$
representation $\pP_J[Q_{ij}]$ is the monomial $Q_{12}^J$, which
defines the (unique) bivalent intertwiner carrying the spin $\f J2$
on both legs 1 and 2.
Finally, in this context, the Pl\"ucker relation on the $Q_{ij}$ variables can truly be interpreted as recoupling relations on intertwiners.
%
%

Before moving on, we would like to comment about the equivalence on using the spinor variables or the $Q_{ij}$ variables or the initial $\lambda,U$ variables. Indeed, at the end of the day, we are considering (anti-holomorphic) functions of the spinors $z_i$ satisfying the closure conditions $\cC$ and also invariant under the $\SU(2)$ transformations that they generate, this defines the manifold $\C^{4N}//\SU(2)= \C^{4N}/\mathrm{SL}(2,\C)$, with dimension:
$$
4N- (3+3).
$$
Let us now compare with the $Q$-matrix defined as $Q=\lambda U\Delta_\eps {}^tU$. As we already said earlier, this defines the manifold $\R_+\times \U(N)/(\SU(2)\times\U(N-2))$ since the expression of $Q$ is invariant under $U\arr UV$ with $V\in \SU(2)\times\U(N-2)$. It is easy to compute the dimension of this manifold:
$$
1+N^2-3-(N-2)^2\,=\,4N-6,
$$
which coincides exactly (as expected!) with the previous dimension of the spinor manifold.

\medskip

In the next section, we will present an alternative construction, which can be considered as ``dual" to the representation defined above. It is based on the coherent states for the oscillators, thus recovering the framework of the $\U(N)$ coherent intertwiner states introduced in \cite{un3} and further developed in \cite{un4}.

\subsection{Intertwiners as Holomorphic Functionals - version 2}

We can also build our Hilbert space of quantum states as a Fock space acting with creation operators on the vacuum state $|0\ra$ of the oscillators. This will be based on the $\SU(2)$ coherent intertwiners and $\U(N)$ coherent states as defined in \cite{un3,un4}.

We start by quantizing the spinors components $z_i^0,\bz_i^0$ and
$z_i^1,\bz_i^1$ satisfying the classical Poisson bracket
\Ref{bracketz} as the creation and annihilation operators of
harmonic oscillators, respectively $a_i,a\dag_i$ and $b_i,b\dag_i$.
We will have these operators acting on the standard coherent states
for quantum oscillators by multiplication by $z$ and derivative with
respect to $z$.

More precisely, let us begin by introducing the basis of $\SU(2)$
coherent intertwiners as defined in \cite{un3,un4} in terms of the
spinors $z_i\in\C^2$ and some extra spin labels $j_i\in\N/2$:
\be
||\{j_i,z_i\}\ra\,\equiv\,
\int_{\SU(2)}dg\,g\vartriangleright
\prod_i\f{(z_i^0a\dag_i+z_i^1b\dag_i)^{2j_i}}{\sqrt{(2j_i)!}}\,|0\ra,
\ee
where $|0\ra$ is the vacuum states of the harmonic oscillators, $a_i\,|0\ra\,=\,b_i\,|0\ra\,=0$. The group-averaging is taken over $\SU(2)$ with its standard action on spinors as $2\times 2$ matrices. This is exactly the $\SU(2)$ transformations generated by $\vJ$ \cite{un3,un4}. We further introduce the $\U(N)$ coherent states\footnotemark:
\be
|J,\{z_i\}\ra
\,\equiv\,
\sum_{\sum_i j_i=J}\f{1}{\sqrt{(2j_i)!}}\, ||\{j_i,z_i\}\ra
\,=\,
\f1{(2J)!}\,\int_{\SU(2)}dg\,g\vartriangleright
\prod_i{\left(\sum_i (z_i^0a\dag_i+z_i^1b\dag_i)\right)^{2J}}\,|0\ra.
\ee
\footnotetext{
As was shown in \cite{un4}, these states are closely related to the coherent state basis for the quantum oscillators:
$$
\sum_{J\in\N} \beta^{2J}\,|J,\{z_i\}\ra
\,=\,
\int dg\,g\vartriangleright
e^{\beta[{\sum_i z_i^0a\dag_i+z_i^1b\dag_i}]}\,|0\ra.
$$
This works because the integral over $\SU(2)$ of odd powers of $a\dag$ and $b\dag$ vanishes.}

Now, we can define our operators $\wM_{ij}$, $\wQ_{ij}$ and $\wbQ_{ij}$ as differential operators in the $z_k$'s acting in the basis $|J,\{z_i\}\ra$ and we can check that they exactly match the action of the operators $E_{ij}$,$F_{ij}$ and $F\dag_{ij}$.

\begin{res}
We define the operators $\wM_{ij}$, $\wQ_{ij}$ and $\wbQ_{ij}$ as differential operators acting on holomorphic functionals $|\vphi\ra \,\equiv\,\int [d^2z]^{2N}\,\vphi(z_k)\,|J,\{z_k\}\ra$~:
\beq
\wM_{ij} &=&
-\left(\f{\pp}{\pp z_i^0}z_j^0+\f{\pp}{\pp z_i^1} z_j^1\right)
=
-\left(z_j^0\f{\pp}{\pp z_i^0}+z_j^1\f{\pp}{\pp z_i^1}\right) -2\delta_{ij} \\
\wQ_{ij} &=&
\f{\pp^2}{\pp z_i^0\pp z_j^1}-\f{\pp^2}{\pp z_i^1\pp z_j^0}.\nn\\
\wbQ_{ij} &=&
z_i^0z_j^1-z_i^1z_j^0 \,=\, \bQ_{ij} \nn
\eeq
These differential operators exactly reproduce the action of  respectively the operators $E_{ij}=a\dag_ia_j+b\dag_ib_j$, $F_{ij}=a_ib_j-a_jb_i$ and $F\dag_{ij}$ on the (coherent) states $|J,\{z_k\}\ra$.

\end{res}

\begin{proof}
We start by computing the action of the operators $E_{ij}$,$F_{ij}$ and $F\dag_{ij}$ on the states $||\{j_k,z_k\}\ra$. In order to do this, we use the definition of those states and simply compute the commutator of the $E_{ij}$,$F_{ij}$ and $F\dag_{ij}$ with the operators $(z_k^0a\dag_k+z_k^1b\dag_k)^{2j_k}$. Then it is straightforward to get:
\beq
E_{ij}\,||\{j_k,z_k\}\ra &=&
\f{\sqrt{2j_j}}{\sqrt{2j_i+1}}\,
\left(z_j^0\f{\pp}{\pp z_i^0}+z_j^1\f{\pp}{\pp z_i^1}\right)\,||\{j_i+\f12,j_j-\f12,j_k,z_k\}\ra,\\
F_{ij}\,||\{j_k,z_k\}\ra &=&
\sqrt{2j_j}\sqrt{2j_i}\,
(z_i^0z_j^1-z_i^1z_j^0)\,||\{j_i-\f12,j_j-\f12,j_k,z_k\}\ra,
\nn\\
F\dag_{ij}\,||\{j_k,z_k\}\ra &=&
\f{1}{\sqrt{(2j_i+1)}\sqrt{2j_i+1}}\,
\left(\f{\pp^2}{\pp z_i^0\pp z_j^1}-\f{\pp^2}{\pp z_i^1\pp z_j^0}\right)
\,||\{j_i+\f12,j_j+\f12,j_k,z_k\}\ra,\nn
\eeq
which allows to obtain the action on the $|J,\{z_k\}\ra$ states:
\beq
E_{ij}\,|J,\{z_k\}\ra &=&
\left(z_j^0\f{\pp}{\pp z_i^0}+z_j^1\f{\pp}{\pp z_i^1}\right)\,|J,\{z_k\}\ra,\\
F_{ij}\,|J,\{z_k\}\ra &=&
(z_i^0z_j^1-z_i^1z_j^0)\,|J-1,\{z_k\}\ra,
\nn\\
F\dag_{ij}\,|J,\{z_k\}\ra &=&
\left(\f{\pp^2}{\pp z_i^0\pp z_j^1}-\f{\pp^2}{\pp z_i^1\pp z_j^0}\right)
\,|J+1,\{z_k\}\ra,\nn
\eeq
These expressions were actually already derived \cite{un4} by other means.
Using these actions of the operators $E_{ij}$,$F_{ij}$ and $F\dag_{ij}$ on the states $||J,\{z_k\}\ra$, we finally derive  their action on states $\int [d^2z]^{2N}\,\vphi(z_k)\,|J,\{z_k\}\ra$ by integration by parts and we recover the expressions given above.
\end{proof}

As in the previous section, we can check that these operators
$\wM_{ij}$, $\wQ_{ij}$ and $\wbQ_{ij}$ satisfy the exact expected
commutation relations and quadratic constraints (when acting on
$\SU(2)$-invariant states), and thus provide a proper quantization
of our classical Poisson structure \Ref{commM}. We notice that it is
the operator $\wbQ_{ij}=F_{ij}$ which now acts as a multiplication
operator while $\wQ_{ij}=F\dag_{ij}$ becomes a derivative operator.
In this sense, we can consider this quantization scheme as ``dual"
to the one presented in the previous section. This comes from
quantizing $z$ as $\pp_{\bz}$ in the previous scheme while
quantizing $\bz$ as $-\pp_z$ in the present scheme based on the
coherent states. For more details on the $\U(N)$ coherent state
basis, the interested reader can refer to \cite{un3,un4}.

\section{Building Holonomies for Loop Gravity}\label{sectionholo}


Up to now, we have described the Hilbert space of a single
intertwiner, corresponding to a single vertex of a spin network
state, in terms of spinors and $\U(N)$ operators. More precisely, we
have described the classical system of $2N$ spinors constrained by
the closure conditions, which is isomorphic to the coset space
$\U(N)/\SU(2)\times\U(N-2)$, and we have explained how its
quantization leads back to the space of $N$-valent intertwiner
states.

In this section, we discuss the generalization of this framework to
whole spin network states for Loop Quantum Gravity. We explain how
to glue intertwiners, or more precisely how to glue these systems of
spinors together along particular graphs. The main result is how to
express holonomies in terms of the spinors. This allows to view spin
network states as functionals of our $Q_{ij}$ variables and fully
reformulate the kinematics of Loop Quantum Gravity in terms of
spinors and the $\U(N)$ operators.

\subsection{Revisiting Spin Network States}


Building on the previous works on the $\U(N)$ framework for
intertwiners \cite{un2,un3,2vertex} and the twistor representation
of twisted geometries for loop gravity \cite{twisted,twistor}, we
would like to give a full representation of the spin network states
in terms of spinors.

Let us start by considering a given oriented graph $\Gamma$, with $E$ edges and $V$ vertices. Let us call $s(e)$ and $t(e)$ respectively the source and target vertices of each edge. Then the Hilbert space of cylindrical functions for Loop Quantum Gravity consists in all functions of $E$ group elements $g_e\in\SU(2)$ which are invariant under the $\SU(2)$-action at each vertex:
\be
\forall h_v\in\SU(2)^{\times V},\quad
\phi(g_e)=\phi(h_{s(e)}g_e h_{t(e)}^{-1}).
\ee
The scalar product between two such functionals is defined by the straightforward integration with respect to the Haar measure on $\SU(2)$:
\be
\la \phi|\tphi\ra
\,=\,
\int_{\SU(2)^{\times E}} [dg_e]\,
\overline{\phi(g_e)}\,\tphi(g_e)\,
\ee
so that the Hilbert space of $\SU(2)$-invariant cylindrical functions on the considered graph $\Gamma$ is
$$
\cH_\Gamma\,\equiv\,L^2(\SU(2)^E/\SU(2)^V).
$$
A basis of this space is given by the spin network states, which are labeled by one $\SU(2)$-representation $j_e$ on each edge $e$ and one intertwiner state $I_v$ on each vertex $v$ of the graph. One goal is to make the link between this and our formalism based on spinors, $Q_{ij}$ variables and $\U(N)$ operators.

\medskip

The first step was already described in \cite{un2}. We consider one
intertwiner state constructed with the $\U(N)$ formalism, and then
we glue them along the edges of the graph. More precisely, we start
with a function $\psi(Q^1,..,Q^V)$ where $Q^v$ is the $N_v\times
N_v$ matrix corresponding to the vertex $v$ where $N_v$ is the
valence of the node $v$. Each of these matrices $Q^v$ is constructed
from a set of spinors $z_{v,e}$ attached to the corresponding vertex
$v$.
These intertwiners are decoupled for now. Following \cite{un2}, we glue them by requiring that they carry the same spin $j_e$ from the point of view of both vertices $s(e)$ and $t(e)$. Since the spin on the leg $e$ of an intertwiner at the vertex $v$ is given by the energy operator $E^v_e=\wM^v_{ee}$ living on that leg, this amounts to imposing the constraint $E^{s(e)}_e-E^{t(e)}_e=0$ on each edge $e$. This {\it matching condition} corresponds to the classical constraint:
\be
M^{s(e)}_{ee} - M^{T(e)}_{ee}
= \la z_{s(e),e}|z_{s(e),e}\ra - \la z_{t(e),e}|z_{t(e),e}\ra
=0,
\ee
which requires that the two spinors $z_{s(e),e}$ and $z_{t(e),e}$ have equal norm.
At the quantum level, this constraint imposes a $\U(1)$-invariance for each edge:
\be
\psi(z_{s(e),e},z_{t(e),e})\,=\,
\psi(e^{i\theta_e}\,z_{s(e),e},e^{-i\theta_e}z_{t(e),e}),\qquad\forall \theta_e\in[0,\pi]\,,
\ee
\be
\psi(Q^{s(e)},Q^{t(e)},Q^v)\,=\,
\psi(e^{-i(\delta_{ie}+\delta_{je})\theta_e}\,Q^{s(e)}_{ij},e^{+i(\delta_{ie}+\delta_{je})\theta_e}\,Q^{t(e)}_{ij},Q^v)
,\qquad\forall \theta_e\in[0,\pi]\,,
\ee
whether we express the wave-functions in terms of the $Q^v_{ij}$
matrix elements or directly in terms of the spinors $z_{v,e}$.
Notice that we multiply the source and target spinors by opposite
phases.

There is two equivalent ways to impose these matching constraints on the wave-functions:
\begin{itemize}

\item Either, we impose $\la z_{s(e),e}|z_{s(e),e}\ra - \la z_{t(e),e}|z_{t(e),e}\ra
=0$ at the classical level on the phase space and consider equivalence classes of spinors under the corresponding $\U(1)^E$ transformations; and then quantize the system by considering (anti-)holomorphic wave-functions on this constrained phase space.

\item
Or quantize the system of intertwiners as we have done up to now without imposing $\la z_{s(e),e}|z_{s(e),e}\ra - \la z_{t(e),e}|z_{t(e),e}\ra
=0$ at the classical level, and then impose the $\U(1)^E$ invariance to the resulting (anti-)holomorphic wave-functions.

\end{itemize}

\begin{conj}
Following this procedure, we consider (anti-)holomorphic wave-functions of the spinors $\psi(z_{s(e),e},z_{t(e),e})$, where all sets of spinors around each vertex $v$ satisfy the closure conditions and invariant under $\SU(2)$ (generated by those same closure conditions), and such that they are invariant under multiplication by a phase on each edge $e$. We conjecture that the $L^2$ space of such functions with respect to the measure \Ref{scalarp} is isomorphic to the Hilbert space $\cH_\Gamma$ of spin network states of the graph $\Gamma$. In more mathematical terms:
\beq
\cH_\Gamma=L^2(\SU(2)^E/\SU(2)^V)
&=&L^2_{holo}\left( \times_v \,\C^{2 N_v}//\SU(2) \right)\,/\U(1)^E \\
&=&L^2_{holo}\big{(} \times_v \,\R^+\times\U(N_v)/(\SU(2)\times\U(N_v-2))\, \big{)}\,/\U(1)^E \nn
\eeq
where the $//\SU(2)$ quotient means that we both impose the closure conditions and the invariance under the $\SU(2)$ transformations that they generate. In this scheme, it is truly the closure conditions at each vertex that induce the $\SU(2)$-gauge invariance of our quantum states.
\end{conj}

A first hint towards establishing this conjecture is a count of the degrees of freedom. Starting by focusing on a given vertex $v$, we are looking at holomorphic functions of $N_v$ spinors  satisfying the closure conditions $\cC$ and invariant under the $\SU(2)$, which gives:
$$
\f12 \left[4 N_v-(3+3)\right]\,=2N_v -3,
$$
taking into account that each spinor counts for 4 real degrees of freedom and the $\f12$-factor accounts for considering only holomorphic functions. We have already commented in section.\ref{antiholo} on the equivalence of counting the number of degrees of freedom defined by the spinor variables or by the $Q$ variables.
We now sum over all vertices $v$ and impose the $\U(1)$ on each edge, which gives:
$$
\sum_v (2N_v -3) - E \,=\, 3E-3V,
$$
since the combinatorics of a graph ensures that the number of edges
can be expressed in terms of the valence of all the nodes as
$2E=\sum_v N_v$.
We compare this to the dimension of the quotient manifold $\SU(2)^E/SU(2)^V$ whose dimension is obviously $3(E-V)$ since $\SU(2)$ has dimension 3 (excluding the ``degenerate" case when $E=V$ which corresponds to a single Wilson loop).

\medskip

The second step towards establishing the correspondence between the
standard formalism of loop (quantum) gravity and our spinor
formulation is provided by the reconstruction of the group element
$g_e$ in terms of the spinors. This was done in \cite{twistor}.

Considering an edge $e$ with the two spinors at each of its end-vertices $z_{s(e),e}$ and $z_{t(e),e}$, there exists a unique $\SU(2)$ group element mapping one onto the other. More precisely:
\be
g_e\,\equiv\,
\f{|z_{s(e),e}]\la z_{t(e),e}|-|z_{s(e),e}\ra [z_{t(e),e}|}
{\sqrt{\la z_{s(e),e}|z_{s(e),e}\ra\la z_{t(e),e}|z_{t(e),e}\ra}}
\ee
is uniquely fixed by the following conditions:
\be
g_e\,\f{|z_{t(e),e}\ra}{\sqrt{\la z_{t(e),e}|z_{t(e),e}\ra}}\,=\,\f{|z_{s(e),e}]}{\sqrt{\la z_{s(e),e}|z_{s(e),e}\ra}},\quad
g_e\,\f{|z_{t(e),e}]}{\sqrt{\la z_{t(e),e}|z_{t(e),e}\ra}}\,=\,-\f{|z_{s(e),e}\ra}{\sqrt{\la z_{s(e),e}|z_{s(e),e}\ra}},\quad
g_e\in\SU(2),
\ee
thus sending the source normalized spinor onto the dual of the target normalized spinor.
Let us point out that if we impose the matching conditions $\la z_{s(e),e}|z_{s(e),e}\ra - \la z_{t(e),e}|z_{t(e),e}\ra =0$ on the spinors, then the norm-factors can be dropped out of the previous equations.
This truly means that the $g_e$'s define the parallel transport of the spinors along the edges of the graph.
This expression $g_e(z_{s(e),e},z_{t(e),e})$ is clearly $\U(1)$-invariant i.e invariant under the simultaneous multiplication by a phase of the two spinors:
$$
z_{s(e),e}\,\arr\, e^{i\theta_e}\,z_{s(e),e},\qquad
z_{t(e),e}\,\arr\, e^{-i\theta_e}\,z_{t(e),e}.
$$
Thus we can consider any function $\phi(g_e)$ as a function $\psi(z_{v,e})$. We would still need to check how the $\SU(2)$ gauge invariant of the  $\phi(g_e)$ functionals are turned into the closure conditions for the wave-functions $\psi(z_{v,e})$.

We postpone a rigorous mathematical study of this issue and the
resulting proof of the conjecture to future investigation
\cite{inprep}. Instead, here, we would like to focus on using this
formula for the $\SU(2)$ group elements in terms of our spinors to
express the holonomy operators of Loop Quantum Gravity in terms of
the $\U(N)$ operators.

\subsection{Reconstructing Holonomies}

The group elements  $g_e(z_{s(e),e},z_{t(e),e})\in\SU(2)$ that we
constructed in the previous section are invariant under $\U(1)$ and
thus commute with the matching conditions $E_e^{s(e)}-E_e^{t(e)}$
ensuring that the energy of the oscillators on the edge $e$ at the
vertex $s(e)$ is the same as at the vertex $t(e)$. However, they are
obviously not invariant under $\SU(2)$ transformation. As well-known
in loop (quantum) gravity, in order to construct $\SU(2)$-invariant
observables, we need to consider the trace of holonomies around
closed loops, i.e the oriented product of group elements $g_e$ along
closed loops $\cL$ on the graph:
\be
G_\cL\,\equiv\,
\overrightarrow{\prod_{e\in\cL}}g_e.
\ee
Let us assume for simplicity's sake that all the edges of the loop
are oriented the same way, so that we can number the edges
$e_1,e_2,..e_n$ with $v_1=t(e_n)=s(e_1)$, $v_2=t(e_1)=s(e_2)$ and so
on. Then, we can write explicitly the holonomy $G_\cL$ in terms of
the spinors:
\be
\tr \,G_\cL\,=\, \tr\,g(e_1)..g(e_n)
\,=\,
\tr\,\f{\prod_i (|z_{v_i,e_i}]\la z_{v_{i+1},e_i}|-|z_{v_{i},e_i}\ra [z_{v_{i+1},e_i}|)}
{\prod_i \sqrt{\la z_{v_i,e_i}| z_{v_{i},e_i}\ra\la z_{v_{i+1},e_i}| z_{v_{i+1},e_i}\ra}}.
\ee
Now, instead of factorizing this expression per edge, let us group the terms per vertex:
\be
\tr\, G_\cL
\,=\,
\sum_{r_i=0,1} (-1)^{\sum_ir_i}
\f{\prod_i \la \varsigma^{r_{i-1}} z_{v_i,e_{i-1}}\, |\, \varsigma^{1-r_i} z_{v_i,e_i}\ra}
{\prod_i \sqrt{\la z_{v_i,e_{i-1}}| z_{v_i,e_{i-1}}\ra\la z_{v_i,e_i}| z_{v_i,e_i}\ra}},
\ee
where the $\varsigma^{r_i}$ records whether we have the term
$|z_{v_i,e_i}]\la z_{v_{i+1},e_i}|$ or $|z_{v_{i},e_i}\ra
[z_{v_{i+1},e_i}|$ on the edge $e_i$. Let us remember that
$\varsigma$ is the (anti-unitary) map sending a spinor $|z\ra$ to
each dual $|z]$.


\begin{figure}[h]
\begin{center}
\includegraphics[height=60mm]{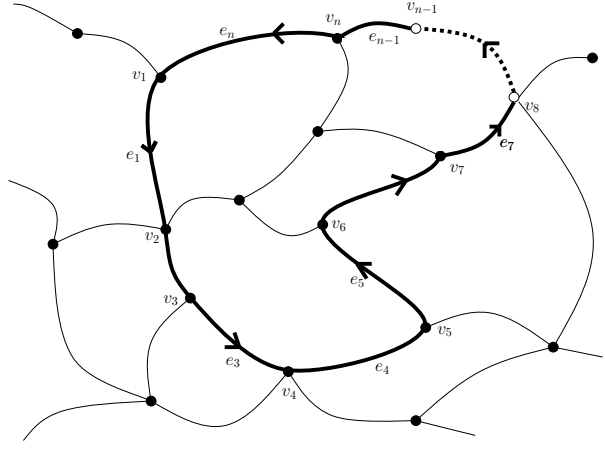}
\caption{The loop $\cL=\{e_1,e_2,..,e_n\}$ on the graph $\Gamma$.}
\end{center}
\end{figure}

Now, depending of the specific values on the $r_i$ parameters, the scalar products at the numerators are given by the matrix elements of $M^i$ or $Q^i$ at the vertex $i$:
\be
\begin{array}{c|c|c}
r_{i-1} & r_i & \la\varsigma^{r_{i-1}} z_{v_i,e_{i-1}}\, |\, \varsigma^{1-r_i} z_{v_i,e_i}\ra \\\hline
0 & 0 & Q^i_{i,i-1} \\
0 & 1 & M^i_{i-1,i} \\
1 & 0 & M^i_{i,i-1} \\
1 & 1 & \bQ^i_{i,i-1}
\end{array}
\ee
Since the matrices $M^i,Q_i,\bQ^i$ are by definition $\SU(2)$-invariant (they commute with the closure conditions), this provides a posteriori check that the holonomy $\tr\, G_\cL$ correctly provides a $\SU(2)$-observables.

Taking into account the various possibilities for the signs $(-1)^{r_i}$, we can write the holonomy in a rather barbaric way:
\be
\tr \,G_\cL\,=\,
\sum_{r_i=0,1} (-1)^{\sum_ir_i}
\f{\prod_i r_{i-1}r_i\bQ^i_{i,i-1}\,+\,
(1-r_{i-1})r_iM^i_{i-1,i}\,+\,
r_{i-1}(1-r_i)M^i_{i,i-1}\,+\,
(1-r_{i-1})(1-r_i)Q^i_{i,i-1}
}{\prod_i \sqrt{\la z_{v_i,e_i}| z_{v_{i+1},e_i}\ra}},
\ee
where actually only one
of the four terms is selected for each set of $\{r_i\}$. To simplify the notations, we call $\cM_\cL^{\{r_i\}}$ each term for a fixed set $\{r_i\}$:
\beq
\cM_\cL^{\{r_i\}}
&\equiv&
\prod_i r_{i-1}r_i\bQ^i_{i,i-1}\,+\,
(1-r_{i-1})r_iM^i_{i-1,i}\,+\,
r_{i-1}(1-r_i)M^i_{i,i-1}\,+\,
(1-r_{i-1})(1-r_i)Q^i_{i,i-1} \\
&=&
\prod_i \la \varsigma^{r_{i-1}} z_{v_i,e_{i-1}}\, |\, \varsigma^{1-r_i} z_{v_i,e_i}\ra.\nn
\eeq
Each of these quantities are still $\SU(2)$-invariant observables and are also invariant under the $\U(1)^E$ transformations generated by the matching conditions. So there are genuine observables on the space of spin networks.

\medskip

After having expressed the holonomy observable in terms of the
spinors and $M,Q,\bQ$ matrices at the classical level, our purpose
is to promote it to a quantum operator and express the holonomy
operator acting on spin network states in terms of the
$\U(N)$-operators $E,F,F\dag$. In order to achieve this, looking at
the vertex $v$ and the pair of edges $e,f$, we simply have to
quantize the matrix elements as:
\beq
M^v_{ef} &\arr& E^v_{ef}, \\
Q^v_{ef} &\arr& F^v_{ef}{}\dag, \nn\\
\bQ^v_{ef} &\arr& F^v_{ef}. \nn
\eeq
Therefore the quantization of the holonomy observable is obvious apart from the factors at the denominator. First, we notice that the norm $\la z_{v,e}| z_{v,e}\ra$ for each edge $e$ attached to $v$ is simply the matrix element $M^v_{ee}$ giving the total energy on the leg $e$ for the intertwiner living at $v$. The natural quantization of these terms is thus $E^v_{ee}$. However, we need to take the inverse square-root of these operators and they do have a 0 eigenvalue. We must also face possible ordering ambiguities because all the $E$ and $F$ and $F\dag$ operators do not commute. In order to decide which ordering is right, we draw inspiration from the direct calculation of the holonomy operator for the 2-vertex graph done in \cite{2vertex} and conjecture the following expression:

\begin{conj}
We can express the holonomy operator around a closed loop $\cL$ (assuming that all the edges are oriented the same way) acting on spin network states as:
\be
\widehat{\tr \,G_\cL}\,=\,
\sum_{r_i=0,1} (-1)^{\sum_ir_i}
\cE\,\hcM_\cL^{\{r_i\}}\,\cE,
\ee
with the operators
$$
\cE\,\equiv\, \f1{\prod_i\sqrt{E_{e_i}+1}}
$$
and
$$
\hcM_\cL^{\{r_i\}}\,\equiv\,{\prod_i r_{i-1}r_iF^i_{i,i-1}\,+\,
(1-r_{i-1})r_iE^i_{i-1,i}\,+\, r_{i-1}(1-r_i)E^i_{i,i-1}\,+\,
(1-r_{i-1})(1-r_i)F^i_{i,i-1}{}\dag\,. }
$$

\end{conj}
First, we have written $E_{e_i}$ without reference to any vertex. This is because spin network states satisfy the matching constraints on all edges $E^{s(e)}_{ee}=E^{t(e)}_{ee}$, therefore we write here $E_{e_i}\equiv E^{v_i}_{e_ie_i}= E^{v_i}_{e_{i-1}e_{i-1}}$. In particular, one can easily check that the operator $\prod_j\hcM_j$  commute with all the matching constraints $E^{v_i}_{e_ie_i}-E^{v_i}_{e_{i-1}e_{i-1}}$.
Second, $E_{e_i}$ is the energy operator of the oscillators living on the edge $e_i$, so it has a positive spectrum $\N$. Thus, the shifted operator $E_{e_i}+1$ is still Hermitian and has a strictly positive spectrum $\N^*=\N\setminus\{0\}$. Therefore, the operator $1/\sqrt{E_{e_i}+1}$ is well-defined.

Finally, we  point out that the operator $\widehat{\tr \,G_\cL}$ defined as above is straightforwardly Hermitian.

\medskip

In order to prove this conjecture, we could do a direct calculation of the action of the holonomy operator, check how it acts on all the intertwiners living at the vertices of the loops $\cL$ and compare with the expression above. We believe that a more indirect check but certainly less painful and more enlightening would be to compute the algebra of our conjectured holonomy operators and compare it to the actual well-known holonomy  algebra. We postpone this study to future investigation \cite{inprep}.

\medskip

We nevertheless check our conjectured formula against the exact expression of the holonomy operators for the 2-vertex graph \cite{2vertex}, and it seems  that we have the exact same expressions apart from the sign factor $(-1)^{r_i}$.
Let us look more carefully at this issue.

The 2-vertex graph consists in two vertices $\alpha$ and $\beta$, linked by $N$ edges all oriented in the same direction from $\alpha$ to $\beta$. We number the edges $i=1..N$. We now have $\U(N)$ operators acting at each vertex, $\Ea_{ij},\Fa_{ij},\Fa_{ij}{}\dag$ and $\Eb_{ij},\Fb_{ij},\Fb_{ij}{}\dag$. Finally, the matching conditions to ensure that we are working with true spin network states are $\Ea_{ii}-\Eb_{ii}=0$ for all edges $i$.


\begin{figure}[h]
\begin{center}
\includegraphics[height=40mm]{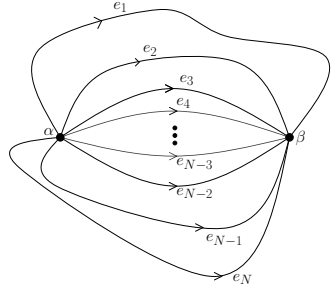}
\caption{The 2-vertex graph with vertices $\alpha$ and $\beta$ and the $N$ edges linking them.}
\end{center}
\end{figure}

Let us look at a basic loop consisting in two edges $(ij)$. Then we apply our conjectured formula to get:
\be
\widehat{\tr \,G_{(ij)}}\,=\,
-\,\f1{\sqrt{E_i+1}\sqrt{E_j+1}}\,
(\Fa_{ij}\Fb_{ij}+\Ea_{ij}\Eb_{ij}+\Ea_{ji}\Eb_{ji}+\Fa_{ij}{}\dag\Fb_{ij}{}\dag)\
\,\f1{\sqrt{E_i+1}\sqrt{E_j+1}}.
\ee
This is the exact same expression as we have derived in the earlier work \cite{2vertex} apart from the global minus sign. This discrepancy is not an issue since it is only due to the difference of orientation. Indeed, our conjecture formula holds for all edges oriented the same way around the loop $\cL$, while the formula derived in \cite{2vertex} assumes that the edges are all oriented from $\alpha$ to $\beta$. There is no problem with changing the orientations in our formula for the holonomy operator above: we multiply by a minus sign for each edge whose orientation we switch.

At the end of the day, the present framework is totally consistent with the full analysis  of spin network states on the 2-vertex graph done in \cite{2vertex}.

\medskip

By expressing the holonomy operator around a closed loop in terms of
the operators $E_{ij}$, $F_{ij}$ and $F\dag_{ij}$ of the $\U(N)$
formalism, we have finally written a proper $\SU(2)$-invariant
operators acting on spin network states and not only on single
intertwiner states as done up to now \cite{un1,un2,un3,un4}. As we
have said earlier, looking carefully at the expression of the
holonomy operator, each term of the sum over $r_i=0,1$ is also
$\SU(2)$-invariant and commutes with the matching conditions.
Moreover, we can forget about the factors  $\cE$ in the denominator,
which comes from properly normalizing the spinors into order to
define the group elements $g_e$. Finally, we are left with the
operators $\hcM_\cL^{\{r_i\}}$ for each set of values $\{r_i\}$,
which we interpret as defining {\it generalized holonomy operators}
in our $\U(N)$ formalism for loop quantum gravity. These operators
are simply constructed as the product of $E$ or $F$ or $F\dag$
operators acting on the vertices around the loop:
\be
\hcM_\cL^{\{r_i\}}\,\equiv\,{\prod_i r_{i-1}r_iF^i_{i,i-1}\,+\,
(1-r_{i-1})r_iE^i_{i-1,i}\,+\,
r_{i-1}(1-r_i)E^i_{i,i-1}\,+\,
(1-r_{i-1})(1-r_i)F^i_{i,i-1}{}\dag
}.
\ee
These are the natural $\SU(2)$-invariant operators acting on spin network states in the $\U(N)$ formalism.  It is easy to see that they shift the spin $j_e$ of the edges around the loop $e\in\cL$ by $\pm\f12$. For instance, for $r_i=0$ for all $i$ around the loop, then:
$$
\hcM_\cL^{\{r_i=0\}}\,=\,
\prod_i F^i_{i,i-1}{}\dag
$$
raises all the spins around the loop by $+\f12$. On the other hand, for $r_i=1$ for all $i$'s, we decrease all the spins by $-\f12$:
$$
\hcM_\cL^{\{r_i=1\}}\,=\,
\prod_i F^i_{i,i-1}{}.
$$
Now, if we put mixed values around the loop $\cL$, the corresponding operator $\hcM_\cL^{\{r_i\}}$ increases the spins of the edges $i$ with $r_i=0$ and decreases the spins on the edges labeled by $r_i=1$.

We can also reconstruct every $\hcM_\cL^{\{r_i\}}$  operator from the holonomy operator $\widehat{\tr \,G_{(ij)}}$ by suitable insertions of the energy operators $E_{e_i}$ in order to select specific $\pm\f12$ shifts for the spins around the loop. This was done explicitly in the case of the 2-vertex graph in \cite{2vertex} and can get straightforwardly generalized to arbitrary graphs and loops.

Finally, the algebra of these generalized holonomy operators  $\hcM_\cL^{\{r_i\}}$ will be investigated elsewhere \cite{inprep}.

%
%
%

\section{Classical Dynamics for Spin Networks}
\label{dynamics}

\subsection{A Classical Action for Spin Networks}

Let us start by summarizing the classical setting for spin network  states on a given graph $\Gamma$. Spin network states are $V$ intertwiner states -one at each vertex $v$- glued together along the edges $e$ so that they satisfy the matching conditions on each edge.
The phase space consists with the spinors $z_{v,e}$ (where $e$ are edges attached to the vertex $v$, i.e such that $v=s(e)$ or $v=t(e)$) which we constrain by the closure conditions $\vcC^v$ at each vertex $v$ and the matching conditions on each edge $e$. The corresponding action reads:
\be
S_0^{\Gamma}[z_{v,e}]
\,=\,
\int dt\,\sum_v
\sum_{e|v\in\pp e} \left(-i\la z_{v,e}|\pp_t z_{v,e}\ra
+\la z_{v,e}|\Lambda_v| z_{v,e}\ra\right)
 +\sum_e \rho_e\left(
 \la z_{s(e),e}|z_{s(e),e}\ra-\la z_{t(e),e}|z_{t(e),e}\ra
 \right),
\ee
where the $2\times 2$ Lagrange multipliers  $\Lambda_v$ satisfying $\tr\,\Lambda_v=0$ impose the closure constraints and the Lagrange multipliers $\rho_e\in\R$ impose the matching conditions. All the constraints are first class, they generate $\SU(2)$ transformations at each vertex and $\U(1)$ transformations on each edge $e$.

We can describe the same system parameterized by $N_v\times N_v$
unitary matrices $U^v$ and the parameters $\lambda_v$. The matrix
elements $U^v_{ef}$ refer to pairs of edges $e,f$ attached to the
vertex $v$. The closure conditions are automatically encoded in the
requirement that the matrices $U^v$ are unitary. We still have to
impose the matching conditions $M^{s(e)}_{ee}-M^{t(e)}_{ee}=0$ on
each edge $e$ where the matrices $M^v=\lambda_v\,U^v\Delta
U^v{}^{-1}$ are functions of both $\lambda_v$ and $U^v$. The action
then reads:
\be
S_0^{\Gamma}[\lambda_v,U^v]\,=\, \int dt \,
\sum_v\left(-i\,\lambda_v \tr\,U^v\Delta \pp_t {U^v}\dag \,-\,\tr
\Theta_v\,(U^v{U^v}\dag-\id) \right) +\sum_e
\rho_e(M^{s(e)}_{ee}-M^{t(e)}_{ee}),
\ee
where the $\rho_e$ impose the matching conditions as before while the $N_v\times N_v$ matrices $\Theta_v$ are the Lagrange multipliers for the unitarity of the matrices $U^v$. Moreover, this action is invariant under the action of $\SU(2)\times\U(N_v-2)$  at every vertex, which reduces the number of degrees of freedom of the matrices $U^v$ to the spinors $z_{v,e}$ which are actually the two first columns of those matrices.

\medskip

This free action describes the classical kinematics of spin networks on the graph $\Gamma$. Now, we would like to add interaction terms and a Hamiltonian to this action in order to define a non-trivial dynamics for the system. Such interaction terms need to be compatible with the closure conditions and the matching conditions i.e be invariant under $\SU(2)$ at each vertex $v$ and $\U(1)$ on each edge. The natural candidates are the generalized holonomy observables $\cM_\cL^{\{r_i\}}$ which we described in the previous section. Our proposal for a classical action for spin networks with non-trivial dynamics is thus:
\be
S_{\gamma_\cL^{\{r_i\}}}^{\Gamma}=S_0^{\Gamma}\,+\,
\int dt\,\sum_{\cL,\{r_i\}} \gamma_\cL^{\{r_i\}}\,\cM_\cL^{\{r_i\}},
\ee
where the $\gamma_\cL^{\{r_i\}}$ are the coupling constants giving
the relative weight of each generalized holonomy in the full
Hamiltonian. Let us point out that the generalized holonomies
$\cM_\cL^{\{r_i\}}$ are a priori not independent from each other. We
postpone the analysis of this issue to future investigation.
Instead, we will study in more detail this classical action
principle in the specific case of the 2-vertex graph.

%
%
%
%
%

\subsection{A Matrix Model for the Dynamics on the 2-Vertex Graph}

Coming back to the 2-vertex graph, we have the two vertices $\alpha,\beta$ linked with $N$ edges. The corresponding classical phase space is parameterized by $2N$ spinors $\za_i$ and $\zb_i$. Then we need to impose the closure constraints on both vertices:
\be
\sum_i |\za_i\ra\la \za_i | = \f12 \sum\la \za_i |\za_i\ra\,\id,\qquad
\sum_i |\zb_i\ra\la \zb_i | = \f12 \sum\la \zb_i |\zb_i\ra\,\id,
\ee
and the matching conditions on all $N$ edges:
\be
\forall i,\quad
\la \za_i |\za_i\ra\,=\, \la \zb_i |\zb_i\ra\,.
\ee
These constraints are not straightforward to solve explicitly, since the $\za_i,\zb_i$ are spinors and not just complex numbers.
We can also write the corresponding action principle in terms of the
unitary matrices $\Ua,\Ub$:
\be
S_0[\Ua,\Ub,\lambda_\alpha,\lambda_\beta]\! \equiv\!\! \int\!\!
dt\left(\!
-i\lambda_\alpha\tr\Ua\Delta\pp_t\Ua{}\dag-i\lambda_\beta\tr\Ub\Delta\pp_t\Ub{}\dag
+\!\sum_i \rho_i\left(
\lambda_\alpha(\Ua\Delta\Ua{}\dag)_{ii}-\lambda_\beta(\Ub\Delta\Ub{}\dag)_{ii}
\right)\!\right),
\ee
where we have left implicit the constraints imposing the unitarity of $\Ua$ and $\Ub$.
It is clear that the matching conditions imply that $\lambda_\alpha=\lambda_\beta$. We can thus slightly simplify this action:
\be
S_0[\Ua,\Ub,\lambda] \,\equiv\, \int dt\,\left(-i\,
\lambda\left[\tr\,\Ua\Delta\pp_t\Ua{}\dag\,+\,\tr\,\Ub\Delta\pp_t\Ub{}\dag\right]
\,+\,\sum_i \rho_i\,\left[
(\Ua\Delta\Ua{}\dag)_{ii}\,-\,(\Ub\Delta\Ub{}\dag)_{ii}
\right]\right)\,.
\ee
Geometrically, $\lambda$ represents the total boundary area of the surface separating the two vertices, while $\Ua$ and $\Ub$ describe the shapes and deformations of the two intertwiners sitting at $\alpha$ and $\beta$.

\medskip

We would like to add some dynamics on this basic setting of the
2-vertex graph. Elementary loops on this very simple graph are made
of two edges $(ij)$. Then given such a loop, we have four
possibilities for our generalized holonomy observables
$\cM_\cL^{\{r_i\}}$. The observable $Q^\alpha_{ij}Q^\beta_{ij}$
corresponds at the quantum level to raising the spins $j_i$ and
$j_j$ on both edges $i,j$. Its conjugate
$\bQ^\alpha_{ij}\bQ^\beta_{ij}$ will decrease the spins $j_i$ and
$j_j$  at the quantum level. The observable
$M^\alpha_{ij}M^\beta_{ij}$ will increase the spin $j_i$ while
decreasing the spin $j_j$. The final possibility is
$M^\alpha_{ji}M^\beta_{ji}=\overline{M^\alpha_{ij}M^\beta_{ij}}$
simply reverses the role of the two edges $i$ and $j$. Finally, our
ansatz for a generic action with a non-trivial dynamics reads:
\be
S[\Ua,\Ub,\lambda]
\,\equiv\,
S_0[\Ua,\Ub,\lambda]
\,+\,\int dt\, \sum_{i,j} \left[
\gamma^+_{ij}\,Q^\alpha_{ij}Q^\beta_{ij} + \gamma^-_{ij}\,\bQ^\alpha_{ij}\bQ^\beta_{ij}
+\gamma^0_{ij}\,M^\alpha_{ij}M^\beta_{ij},
\right],
\label{genericaction}
\ee
where the $\gamma$'s are coupling constants and where we remind that
the matrices $M$ and $Q$ are defined as $M=\lambda\,U\Delta U\dag$
and $Q=\lambda\,U\Delta_\eps {}^tU$.
Further requiring that the new interaction terms defining the action's Hamiltonian be real, we need to impose further condition on the coupling constants:
$$
\gamma^- =\overline{\gamma^+},\qquad
\gamma^0=(\gamma^0)\dag.
$$

\medskip

In the previous work on spin networks on the 2-vertex graph
\cite{2vertex}, it was discussed to introduce an extra $\U(N)$
symmetry and further require that the dynamics of the system be
invariant under that symmetry. This was then interpreted as imposing
isotropy on the model. These coupled $\U(N)$ transformations act on
the matrices at both vertices $\alpha$ and $\beta$:
\be
\left|
\begin{array}{l}
\Ua\,\arr\, V\Ua \\
\Ub\,\arr\, \bar{V}\Ub
\end{array}
\right.
\qquad
\textrm{for}\quad
V\in\U(N).
\ee
These $\U(N)$ transformations are generated by
$\wM^\alpha_{ij}-\wM^\beta_{ji}$ and they reduce to the $\U(1)^N$
transformations generated by the matching constraints in the case
that $V$ is a unitary diagonal matrix.

The kinematical terms in $\tr\,\Ua\Delta\pp_t\Ua{}\dag$ and
$\tr\,\Ub\Delta\pp_t\Ub{}\dag$ are obviously  invariant under such
transformations. The interaction terms also need to be
$\U(N)$-invariant. It is easy to check that this leaves only three
possible $\U(N)$-invariant terms made from all the generalized
holonomy observables in the equation \ref{genericaction}:
\be
\gamma^+\,\tr\, Q^\alpha Q^\beta + \gamma^-\,\tr\,\bQ^\alpha\bQ^\beta
+\gamma^0\,\tr\,M^\alpha {}^tM^\beta\,,
\ee
where we remind that ${}^tM=\bar{M}$ and ${}^tQ=-Q$. The requirement to keep the Hamiltonian real imposes as above that $\gamma^-=\overline{\gamma^+}$ and $\gamma^0\in\R$.

Finally, we need to deal with the matching conditions $M^\alpha_{ii}-M^\beta_{ii}=0$. Imposing the invariance under the coupled $\U(N)$-transformations implies imposing the full equality between the matrices $M^\alpha$ and ${}^tM^\beta$, and not only the equality of their matrix elements on the diagonal. This is a very strong condition, which relates the unitary matrices $\Ua$ and $\Ub$ to each other:
\be
M^\alpha={}^tM^\beta
\,\Leftrightarrow\,
\Ua\Delta \Ua{}\dag=\overline{\Ub}\Delta \overline{\Ub}{}\dag
\qquad
\Rightarrow\quad
\Ua\,=\,e^{i\phi} \,\overline{\Ub},
\ee
where $\phi$ is an arbitrary phase factor and the matrices $\Ua$ and $\Ub$ are defined up to $\SU(2)\times\U(N-2)$ transformations. This means that the spinors $\za$ and $\zb$ are also equal up to a global phase:
\be
\bar{z}^{(\alpha)}_i\,=\,e^{i\phi}\,z^{(\beta)}_i,
\ee
which obviously solve the matching conditions. This phase $e^{i\phi}$ actually defines the $\SU(2)$ holonomy living on the edges between the two vertices.

Simply renaming the matrix $\Ub\equiv U$, we can re-express the
action for this $\U(N)$-invariant sector in terms of $\lambda$, the
phase $\phi$ and the matrix $U$. Actually it turns out that the
unitary matrix $U$ completely drops out and we are left with the two
conjugated dynamical variables $\lambda$ and $\phi$:
\be
S_{inv}[\lambda,\phi] \,=\, -2 \int dt\,\left(\lambda \pp_t \phi
-\lambda^2\left(\gamma^0-\gamma^+e^{2i\phi}
-\gamma^-e^{-2i\phi}\right)\right), \label{invaction}
\ee
with the Hamiltonian $H=\lambda^2(\gamma^0-2\gamma\cos(2\phi))$.
This is an elementary action with simple  equations of motion for
the couple of variables $(\phi,\lambda)$. For the sake of
simplicity, we will take $\gamma^+=\gamma^-=\gamma\,\in\R$. Then we
obtain the following equations of motion:
\beq
\pp_t \phi &=& 2\lambda\,(\gamma^0-2\gamma\,\cos\,2\phi)\,, \\
\pp_t \lambda &=& -4\gamma\lambda^2\,\sin\,2\phi\,. \nn
\eeq
We easily identify two obvious classical solutions. First, $\lambda=0$ (with $\phi$ constant and arbitrary) is the trivial solution. It has no evolution and corresponds to a vanishing total area. Second, we have the case where $\phi$ is constant, but $\lambda$ does not vanish. In this case, we get:
$$
\cos\,2\phi\,=\,\f{\gamma^0}{2\gamma},\qquad
\lambda=\f1{(4\gamma\,\sin\,2\phi)\,t+ kk}\,,
$$
where $kk$ is a constant of integration. This solution exists iff
$|\gamma^0|\le 2\gamma$. Finally, we can try to solve the equations
of motion more generally. We express $\lambda$ in terms of
$\pp_t\phi$ from the first equation, which we plug back into the
second equation in order to finally obtain a differential equation
on $\phi$ only:
\beq
\lambda&=&\frac{\partial_t\phi}{2(\gamma^0-2\gamma\cos(2\phi))}\,,\nn\\
(\gamma^0-2\gamma\cos(2\phi))\partial_{t}^2\phi&=&2\gamma\sin(2\phi)(\partial_t\phi)^2\,.\nn
\eeq
Unfortunately, we haven't been able to solve this differential equation explicitly.

On the other hand, we would like to propose an alternative Hamiltonian, who leads to simpler equations of motion which we are able to solve exactly.
Following what has been done in the quantum 2-vertex model presented in \cite{2vertex}, we introduce a renormalized Hamiltonian:
\be
\hh\,\equiv\,
\f1\lambda\,H=\lambda(\gamma^0-2\gamma\cos(2\phi)).
\ee
This renormalized Hamiltonian is still $\SU(2)$ and $\U(N)$ invariant, and is related to the generalized holonomy observables through a factor $\f1\lambda$.
Using this new Hamiltonian $\hh$, the equations of motion actually simplify to
\beq
\partial_t\phi&=&\gamma^0-2\gamma\cos(2\phi),\\
\partial_t\lambda&=&-4\gamma\lambda\sin(2\phi)\,.\nn
\eeq
As in the quantum case \cite{2vertex}, the properties of the
renormalized Hamiltonian $\hh$ are much more straightforward than
the original Hamiltonian $H$. In fact, it is possible to solve
exactly these differential equations.
We solve for $\phi(t)$ analytically. Then, once we have the solution
for $\phi$, we can show that the following expression for $\lambda$
in terms of $\phi$ solves the equations of motions:
\be
\label{conicparam} \lambda=\frac{\eps}{\gamma^0-2\gamma\cos(2\phi)}\,,
\ee
where $\eps=\pm$ is a global sign. Let us point out that the equation of motion for $\lambda$ only determines it up a global numerical factor. Then we should remember that $\lambda$ is the total area and we always constrain it to be positive.

Now, we present the solutions for $\phi(t)$ (we have chosen the most
convenient constants of integration due to the fact that this
constants are just translations in the temporal variable), depending
on the different values for the parameters $\gamma^0$ and $\gamma$:
\begin{subequations}
\label{regions}
\begin{eqnarray}
&&\mathbf{Elliptic\,\,\, region\,\,}
(|\gamma^0|>2|\gamma|):\hspace{1.16cm}
\phi(t)=-\arctan\left(\frac{(2\gamma-\gamma^0)\tan\left(t\sqrt{(\gamma^0)^2-4\gamma^2}\right)}
{\sqrt{(\gamma^0)^2-4\gamma^2}}\right)\,,\\
&&\mathbf{Hyperbolic\,\,\,
region\,\,}(|\gamma^0|<2|\gamma|):\hspace{0.5cm}
\phi(t)=-\arctan\left(\frac
{\sqrt{4\gamma^2-(\gamma^0)^2}}{(2\gamma+\gamma^0)\tanh\left(t\sqrt{4\gamma^2-(\gamma^0)^2}\right)}\right)\,,\\
&&\mathbf{Parabolic\,\, region\,I\,\,}
(\gamma^0=2\gamma):\hspace{1.05cm} \phi(t)=-\arctan\left(\frac{1}
{4\gamma t}\right)\,,\\
&&\mathbf{Parabolic\,\, region\,II\,\,}
(\gamma^0=-2\gamma):\hspace{0.6cm} \phi(t)=-\arctan\left(4\gamma
t\right)\,.
\end{eqnarray}
\end{subequations}

\begin{figure}[h]
\begin{center}
\includegraphics[height=100mm]{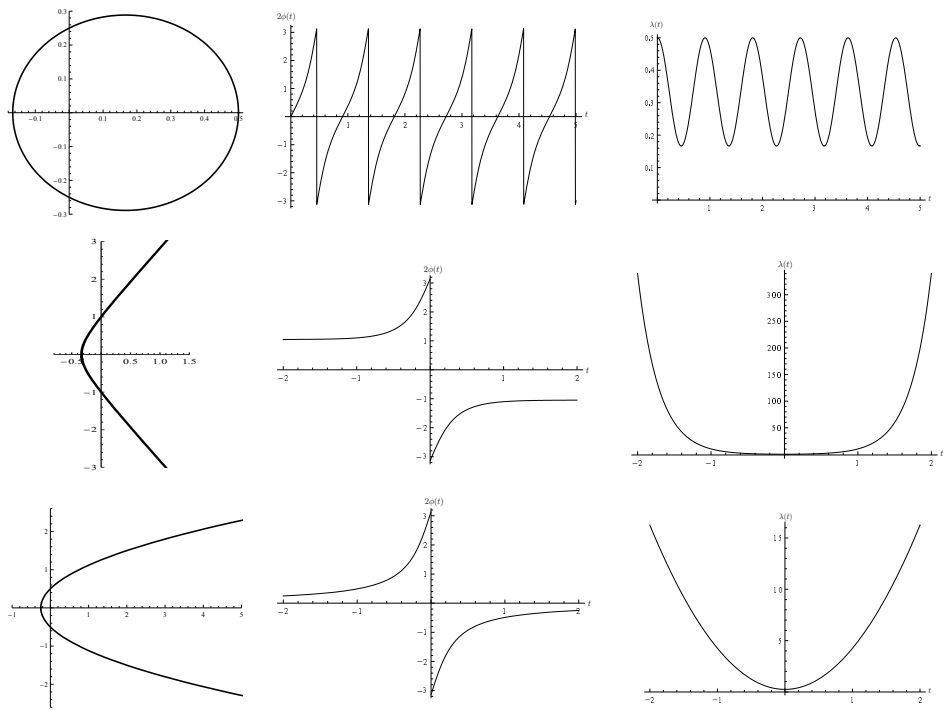}
\caption{We plot the behavior of $\phi(t)$ and
$\lambda(t)$ (given by the equations \ref{conicparam} and
\ref{regions}) in the three different regimes for $\gamma=1$ and respectively $\gamma^0=4$ (elliptic regime), $\gamma^0=1$ (hyperbolic regime) and finally  $\gamma^0=2$ (parabolic regime). In the first column, we give the polar plots constructed by taking as polar
coordinates $(2 \phi,\lambda(\phi))$. The second column gives for
$\phi(t)$ and the third one  $\lambda(t)$. We observe in those
plots the periodical behavior of $\lambda$ (interpreted as the total
area of the model) as a function of time in the elliptic case and
a behavior  analogous to a cosmological big bounce in the other two cases.}
\label{conics}
\end{center}
\end{figure}

Let us give a brief description of these solutions. First, to derive
$\lambda(t)$ from those solutions for $\phi(t)$, we compute $\cos
2\phi\,=\,(1-\tan^2\phi)/(1+\tan^2\phi)$ and we plug it in the
expression \Ref{conicparam} above for $\lambda$ in terms of
$\cos2\phi$.
Then, the two cases I and II of the parabolic regime are very similar. In both cases, we get the same solution $\lambda(t)$ by taking $\eps=+$ in case I and $\eps=-$ in case II. Switching the sign $\eps$ in the two cases allows to keep a positive solution $\lambda(t)\ge 0$.
%
%
We have named the different regions with the name of the conics
because, indeed, the equation \ref{conicparam} is the equation for a
conic with radial coordinate given by $\lambda$ and polar coordinate
$2\phi$, as we can appreciate in the figure \ref{conics}. Then, the
different values for the ratio $2\gamma/\gamma^0$ represent the
different eccentricities corresponding to each of the conics. In the
elliptic case, when $|\gamma^0|>2|\gamma|$, we have a system in
which the area $\lambda$ has an oscillatory behavior. In the other
two regimes (hyperbolic with $|\gamma^0|<2|\gamma|$ and parabolic
with $|\gamma^0|=2|\gamma|$), the area shrinks under evolution,
reaches a minimum value and then increases until infinity. As it was
pointed out in \cite{2vertex}, the quantum Hamiltonian of this
$2$-vertex model is mathematically (and physically) analogous to the gravitational
part of the Hamiltonian in loop quantum cosmology (LQC). Following this
analogy, we can interpret the results we obtained here as the
classical model for the quantum big bounce found in LQC.
Nevertheless, further investigation is needed in order to achieve a
full understanding of the relation with the results derived in the
LQC framework.

\medskip

At this point, it would be enlightening to compare the phase space
$(\phi,\lambda)$ and the dynamics defined by the action
\Ref{invaction} above with the classical setting of loop cosmology
(see e.g. \cite{lqc}). In this sense, the framework presented here
opens at least two interesting lines of research upon understanding
the precise and explicit links between the 2-vertex framework and
the loop cosmology.

First, we should go beyond the $\U(N)$-invariant sector. Indeed, the
action \Ref{genericaction} defines the full classical kinematics and
dynamics of spin network states on the 2-vertex graph. It is a
non-trivial matrix model defined in terms of the unitary matrices
$\Ua$ and $\Ub$ and with quartic interaction terms. Moreover, even
if we still choose a $\U(N)$-invariant Hamiltonian of the the type
$\gamma^+\,\tr\, Q^\alpha Q^\beta +
\gamma^-\,\tr\,\bQ^\alpha\bQ^\beta +\gamma^0\,\tr\,M^\alpha
{}^tM^\beta$, this will nevertheless induce non-trivial dynamics for
the matrices $\Ua$ and $\Ub$. It would be very interesting what kind
of anisotropy does our model describe in the context of loop
cosmology.

Second, our classical phase space expressed in terms of spinors or
equivalently in terms of unitary matrices admit a straightforward
quantization in terms of $\U(N)$ representations. This quantization
scheme should be compared to the quantization procedure of loop
quantum cosmology. This would help understanding the explicit
relation between loop quantum cosmology and the full theory of loop
quantum gravity.

Finally, a constant issue would be to couple matter degrees of
freedom to our model, both in the specific case of the 2-vertex
model and in the general case of spin network states on an arbitrary
graph. The advantage of our approach here is that we can do it at
the classical level in our spinor phase space before quantizing.



\section*{Conclusion}

The $\U(N)$ framework introduced in \cite{un1,un2,un3,un4} provides us with a new
interesting way to describe the space of intertwiners for loop
quantum gravity. Using this framework, it has been shown that it is
possible to tackle some of the important issues in LQG, such as the construction of coherent
states \cite{un3}, dynamics \cite{2vertex} or the simplicity
constraints for spinfoam models \cite{un4}.
Our motivation for the present paper was to define the classical phase space (using spinors) underlying the $\U(N)$ framework and to introduce the corresponding classical action principle.

As it was suggested in \cite{2vertex},  we explored in this paper
the idea of considering the operators $E_{ij}$ and $F_{ij}$ coming
from the quantization of matrix elements of a hermitian matrix $M$
and an antisymmetric matrix $Q$ that satisfy the same quadratic
constraints as the operators themselves (up to quantum ordering
terms). We gave the explicit expression for this matrices in terms
of elements of a unitary matrix $U$ and the parameter $\lambda$
corresponding to the trace of the matrix $M$ (whose quantum
analog is the total area operator $E$). This allowed us to write these matrices
in terms of spinors defined from the matrix elements of the unitary
matrix. This show how the so-called closure conditions for
the spinors, introduced in \cite{un3}, come from the unitarity requirement in the construction of our matrices $M$ and $Q$.
We then described the phase space in terms of the spinors,
introduced the corresponding Poisson brackets and showed that the
closure constraints generate the $\SU(2)$ action relevant to
defining intertwiners states. We further computed the Poisson
brackets of the matrix elements $M_{ij}$ and $Q_{ij}$ and compared
them to the commutator algebra of the corresponding quantum
operators $E_{ij}$ and $F_{ij}$. Finally, we proposed an action from
which we can derive the whole spinor phase space structure. We wrote
it alternatively in terms of the spinors or in terms of the unitary
matrix $U$ and the classical boundary area $\lambda$. This way,  we
have reformulated the classical setting of a single $\SU(2)$
intertwiner as a unitary matrix model (as it was already suggested
in the pioneer work \cite{un1}). Using this action principle, we
finally  showed that there is no non-trivial dynamics for a single
intertwiner.

We moved on to the quantum level and showed how to perform the
quantization of the  classical spinor phase space in order to obtain
the Hilbert space of intertwiners in terms of holomorphic (or
alternatively anti-holomorphic) wave-functions.
Having explored in detail the framework for a single intertwiner
both at the classical and quantum level, we studied the gluing of
those intertwiners and showed to define loop quantum gravity's
spin-network states over an arbitrary graph as (anti-)holomorphic
wave functions of spinors (appropriately constrained). We have
postponed a more rigorous proof of the equivalence of the standard
loop quantum gravity framework to our new $\U(N)$/spinor framework
to future investigation \cite{inprep}.
Then, making use of the expression for $\SU(2)$ group elements in
terms of spinors given in \cite{twistor}, we constructed the
expression for loop gravity's holonomy observables in terms of
spinors attached to each of the vertices of the graph and, finally,
in terms of the elements of the matrices $M$ and $Q$. We discuss the
quantization of this expression and express the holonomy operator at
the quantum level in terms of  the $E_{ij}$ and $F_{ij}$ operators
of the $\U(N)$ framework for intertwiners.
We checked this
formula against the formula derived for the (loop) quantum gravity 2-vertex model previously introduced and discussed by some of the authors \cite{2vertex}.

Finally, we wrote an action for the classical setting of
spin-network states on a general graph and we applied it to the
special case of the simple graph with two vertices. Choosing a
specific form for the interaction term, written in terms of the
(generalized) holonomy observables, we obtained the classical
description for the $\U(N)$ invariant Hamiltonian for the 2-vertex
model considered in \cite{2vertex}.

To summarize, we have proposed here a classical setting whose
quantization in terms of (anti-)holomorphic functionals over constrained spinors describes
intertwiners and spin network states. Furthermore, we proposed
a classical action principle encoding  the whole corresponding kinematical structure and possible dynamics for spin network states.

\section*{Acknowledgments}

This work was in part supported by the Spanish MICINN research
grants FIS2008-01980 and FIS2009-11893. IG is supported by the
Department of Education of the Basque Government under the
``Formaci\'{o}n de Investigadores'' program.

EL acknowledged support from the European Science Foundation (ESF)
through the Short Visit Travel Grant 3595 and from the Programme
Blanc LQG-09 from the ANR (France).


\end{document}